\documentclass[twocolumn,showpacs,nofootinbib,aps,prd,article,article]{revtex4}
\usepackage{soul,graphicx,dcolumn,bm,amsfonts,amsmath,color,xcolor,amsthm,cancel}
\usepackage[dvipdfm, colorlinks=true, citecolor=blue, linkcolor=blue,urlcolor=blue]{hyperref}
\usepackage{slashed}
\usepackage{ulem}

\allowdisplaybreaks

\begin{document}
\title{Vector mesons leading-twist longitudinal distribution amplitudes and related semi-leptonic decays within QCD sum rules}
\author{Wan-Bing Luo}
\address{School of Physics and Mechatronic Engineering, Guizhou Minzu University, Guiyang 550025, P.R.China}
\author{Ru-Meng Pan}
\address{School of Physics and Mechatronic Engineering, Guizhou Minzu University, Guiyang 550025, P.R.China}
\author{Ya-Xiong Wang}
\address{School of Physics and Mechatronic Engineering, Guizhou Minzu University, Guiyang 550025, P.R.China}
\author{Tao Zhong\footnote{Corresponding author}}
\email{zhongtao@gzmu.edu.cn}
\address{School of Physics and Mechatronic Engineering, Guizhou Minzu University, Guiyang 550025, P.R.China}

\date{\today}

\begin{abstract}
In this work, we focus on the light vector meson leading-twist longitudinal distribution amplitudes (DAs) $\phi^\parallel_{2;V}(x,\mu)$ with $V = \rho, K^\ast, \phi$. A new scheme of QCD sum rule research with respect to DA suggested in 2021 by us is adopted. With an improved sum rule formula, the $\xi$-moments $\langle\xi^n\rangle_{2;V}^\parallel$ up to tenth order are calculated. In which, $\langle\xi^2\rangle^\parallel_{2;\rho}=0.225^{+0.013}_{-0.012}$, $\langle\xi^1\rangle^\parallel_{2;K^\ast}=-0.0228^{+0.0042}_{-0.0040}$, $\langle\xi^2\rangle^\parallel_{2;K^\ast}=0.217^{+0.007}_{-0.007}$, $\langle\xi^2\rangle^\parallel_{2;\phi}=0.209^{+0.020}_{-0.020}$, and the corresponding Gegenbauer moments $a^{2;\parallel}_{2;\rho}=0.074^{+0.039}_{-0.036}$, $a^{1;\parallel}_{2;K^\ast}=-0.038^{+0.007}_{-0.007}$, $a^{2;\parallel}_{2;K^\ast}=0.050^{+0.020}_{-0.019}$, $a^{2;\parallel}_{2;\phi}=0.027^{+0.058}_{-0.058}$ at the scale $\mu = 1~{\rm GeV}$, respectively. By fitting those $\langle\xi^n\rangle^\parallel_{2;V}(n = 1,2,\cdots,10)$ with the least squares method, the behaviors of leading-twist longitudinal DAs for $\rho, K^\ast, \phi$ are determined. In addition, we also analyze the impact of incorporating different numbers of $\xi$-moments into the fitting on the stability and reliability of the obtained DAs. Be different from the pion leading-twist DA, we find that the higher-order moments do not significantly change the reconstructed shapes of $\phi^\parallel_{2;V}(x,\mu)$. Further, we recalculate the transition form factors and branching ratios of the $D\to(\rho,K^\ast)$, $D_s\to\phi$ semi-leptonic decay processes.
\end{abstract}


\maketitle

\section{Introduction}
As a universal non-perturbative parameter, distribution amplitude (DA) stands for the momentum fraction distribution of partons in hadrons. Based on the QCD factorization theory, DA enters the exclusive process with large momentum transition through vacuum-hadron matrix element parametrization. More specifically, physical quantities can usually be represented in the form of convolutions with respect to DAs arranged by their different twist structures. In which the leading-twist DA dominates due to higher power suppressed contributions from higher twists. Therefore, the meson leading-twist DA is the key input parameter and main error source in the theoretical research of related exclusive processes and further combining experimental measurements to test standard model and explore new physics. In this work, we focus on the leading-twist longitudinal DAs $\phi^\parallel_{2;V}(x,\mu)$ of light vector mesons $V = \rho$, $K^\ast$ and $\phi$.

The DAs $\phi^\parallel_{2;V}(x,\mu)$ can be studied by various theoretical methods, such as QCD sum rules (QCD SRs)~\cite{Ball:1998sk, Ball:2004rg, Pimikov:2013usa, Bakulev:1998pf, Ball:1996tb, Stefanis:2015qha, Ball:2007zt, Ball:2003sc,Lin:2025cmn}, lattice QCD (LQCD)~\cite{Arthur:2010xf, Boyle:2008nj, Braun:2016wnx, Hua:2020gnw, QCDSF-UKQCD:2007xbk}, Dyson-Schwinger equation (DSE)~\cite{Maris:2003vk, Gao:2014bca, Lu:2021sgg, Xu:2025hjf}, anti-de Sitter/quantum chromodynamics (AdS/QCD) holographic duality method~\cite{Ahmady:2012dy, Ahmady:2013cva, Forshaw:2012mb,Forshaw:2012im, Ahmady:2013cga}, bethe-salpeter wave functions (BSWF)~\cite{Serna:2022yfp}, light-front quark model (LFQM)~\cite{Choi:2007yu, Choi:2013mda, Arifi:2025olq, Dhiman:2019ddr, Puhan:2025ujg}, light-cone quark model (LCQM)~\cite{Ji:1992yf}, light-front hooft equation (LFH)~\cite{Gurjar:2024wpq} and large momentum effective theory (LMET)\cite{Xu:2018mpf}, or combining phenomenological models with experimental data~\cite{Forshaw:2010py}, etc. It is worth noting that existing results show that different theoretical methods give significantly different predictions for the moments and the behaviors of the light vector meson DAs. For example, the DSE~\cite{Xu:2025hjf} determines the DA behaviors by calculating the first to eighth order Mellin moments of the light vector mesons. The DAs given by it are significantly wider than other theoretical predictions in the endpoint region. The AdS/QCD~\cite{Forshaw:2012im} gives the second $\xi$-moment of the $\rho$ meson as 0.228. Other theoretical methods give predictions ranging from 0.193 to 0.263~\cite{Arthur:2010xf,Serna:2022yfp, Choi:2007yu, Arifi:2025olq, Pimikov:2013usa, Gurjar:2024wpq}. The $\phi$ meson DAs behavior obtained from the BSWF~\cite{Serna:2022yfp} is significantly higher in the intermediate momentum fraction region ($(x\sim0.5)$ than that by DSE~\cite{Gao:2014bca}, LQCD~\cite{Hua:2020gnw}, and asymptotic form (AF). In addition, the LFQM~\cite{Choi:2013mda} predictions for the first two $\xi$-moments of the $\rho$ and $K^\ast$ are lower than those from LQCD~\cite{Arthur:2010xf, Boyle:2008nj, Braun:2016wnx, Hua:2020gnw}, QCD SRs~\cite{Ball:2004rg, Pimikov:2013usa, Bakulev:1998pf, Ball:1996tb, Stefanis:2015qha, Ball:2007zt, Ball:2003sc}, DSE~\cite{Gao:2014bca}, and LFH~\cite{Gurjar:2024wpq}. These differences arise from distinct treatments of non-perturbative QCD dynamics. Among the many theoretical methods for studying meson DA, the QCD SRs in the framework of the background field theory (BFTSRs) provides an effective way to calculate the $\xi$-moments~\cite{Zhong:2014jla, Zhong:2021epq}. The background field theory (BFT) gives a field-theoretic description for the basic assumption of QCD SRs, i.e., the non-vanishing vacuum condensates (e.g.~$\langle\bar q q\rangle$, $\langle \alpha_sG^2\rangle$). The basic idea of BFT is to decompose the quark and gluon fields in the QCD Lagrangian or Green function into the classical background fields that satisfy the theoretical equations of motion, and the quantum fields that describe the quantum fluctuations on them. In the calculation of operator product expansion, the vacuum matrix elements of the background fields are naturally parameterized as non-perturbative vacuum condensates, while the quantum fields are calculated perturbatively to obtain the Wilson coefficients.

In principle, more higher-order moments need to be introduced to provide stricter constraints on the DA behaviors. However, the traditional QCD SRs and LQCD research schemes are difficult to give reliable high-order Gegenbauer moments. In our previous work on the pionic leading-twist DA~\cite{Zhong:2021epq}, a new QCD SRs research scheme for meson DA is suggested. Based on the fact that the sum rule of zero-order $\xi$-moment cannot be normalized in the whole Borel parameter region due to the absence of high-dimensional vacuum condensate terms, a new sum rule formula of $n{\rm th}$ $\xi$-moment is proposed to significantly improve its accuracy. By directly fitting the those obtained values of $\xi$-moments with the least squares method, the behavior of DA can be determined. The advantage of this scheme is that it avoids the unreliable higher-order Gegenbauer moments (see the relevant discussion in Ref.~\cite{Zhong:2021epq}) and uses the $\xi$-moments to directly constrain the DA behavior, which allows the provision of a sufficient number of constraints, so as to ensure the reliability and stability of the obtained DA. Therefore, this scheme will be used for the research on DAs $\phi_{2;V}(x,\mu)$ within BFTSRs in this work. Meanwhile, considering the difference between vector mesons and pion, we will analyze in detail the influence of the number of $\xi$-moments involved in fitting on the obtained behaviors of $\phi_{2;V}(x,\mu)$.

Furthermore, we will apply these obtained DAs to calculate the transition form factor (TFFs) and branching ratio for the semi-leptonic decays $D_{(s)}\to V$ with the light-cone QCD sum rules (LCSRs) method. Experimentally, the CLEO, BaBar, and BESIII collaborations have measured the semi-leptonic decays $D_{(s)}\to V$ ~\cite{BESIII:2024qnx, BESIII:2024lxg, BESIII:2023opt, CLEO:2011ab, BESIII:2021pvy, CLEO:2005rxg, BaBar:2008gpr, BESIII:2017ikf, CLEO:2010enr, CLEO:1994msc, BESIII:2018qmf, BESIII:2025fso, Hietala:2015jqa}. Specifically, the BESIII collaboration used the $e^+e^-$ annihilation data collected at $\sqrt s=3.773 \rm{GeV}$ with an integrated luminosity of 7.93 fb$^{-1}$ to study the semi-leptonic decay $D^0\to K^-\pi^0\mu^+\nu_\mu$ for the first time in 2025. They extracted the branching fraction and the TFFs of $D^0\to K^\ast(892)^-\mu^+\nu_\mu$. The precision of the branching fraction was about five times better than the previous world average, and no evidence of lepton flavor universality violation was found~\cite{BESIII:2024qnx}. Meanwhile, the BESIII collaboration reported precise measurements of the branching fractions and TFFs for the semi-leptonic decays $D^0\to\rho^- e^+\nu_e$ and $D_s^+\to\phi\mu^+\nu_{\mu}$ in 2023~\cite{BESIII:2023opt} and 2024~\cite{BESIII:2024lxg}, respectively. Notably, experimental measurements have become increasingly precise. Obvious deviations between some theoretical predictions and experimental data have emerged. The Particle Data Group (PDG) world average of the branching fraction for $D^+\to \bar{K}^{\ast0} e^+\nu_e$ is $(5.40\pm 0.10)\%$~\cite{ParticleDataGroup:2024cfk}, while different theoretical models give predictions ranging $(5.28-8.35)\%$~\cite{APE:1994kxx, Fu:2020vqd, Wu:2006rd, Ivanov:2019nqd, Soni:2017eug, Melikhov:2000yu, Sekihara:2015iha, Fajfer:2005ug}. The BESIII measurements of the TFFs ratios $r_V$ and $r_2$ in $D^0\to K^\ast(892)^-\mu^+\nu_\mu$ deviate from the central values of the combined heavy meson and chiral theory (HM$\chi$T), the covariant light-front quark model (CLFQM), and covariant confining quark model (CCQM), the heavy quark effective theory (HQEFT), the relativistic quark model (RQM) by more than three standard deviations ~\cite{BESIII:2024qnx}. The BESIII measurement of the branching fraction for $D^0\to\rho^- e^+\nu_e$ is $(1.439 \pm 0.033 \pm 0.027) \times 10^{-3}$~\cite{BESIII:2024lxg} also differs from the theoretical predictions in Refs.~\cite{Cheng:2017pcq, Wu:2006rd, Leng:2020fei, Faustov:2019mqr, Scora:1995ty,Fajfer:2005ug, Wang:2002zba, Soni:2018adu, Sekihara:2015iha}.

In theory, the core of the study of $D_{(s)}\to V$ semi-leptonic decays lies in the accurate calculation of the TFFs. The transition matrix element can be decomposed into four TFFs according to the Lorentz structure~\cite{Du:2003ja}. These TFFs encode the complete hadronic information from the non-perturbative QCD effects at low $q^2$ to the bound-state dynamics at high $q^2$. Different $q^2$ regions involve different physical mechanisms. The calculation of the TFFs by different theoretical methods is also applicable to different $q^2$ regions. For example, LQCD~\cite{Flynn:1997ca, Lubicz:1991bi, Bernard:1991bz, APE:1994kxx, Lubicz:1990pi, Gill:2001jp, Donald:2011ff, Donald:2013pea} and HM$\chi$T~\cite{Fajfer:2005ug} are applicable in the large $q^2$ region, while the three-point sum rules (3PSRs)~\cite{Du:2003ja,Ball:1993tp} and LCSRs~\cite{Wu:2006rd, Aliev:2004vf} are suitable for the low and intermediate $q^2$ regions. In contrast, CLFQM~\cite{Galkin:2025emi} and RQM~\cite{Faustov:2019mqr} are both based on meson wave function models and can directly provide TFFs predictions in the entire $q^2$ region. However, their results depend more strongly on the model parameters. In addition, the LFQM~\cite{Wang:2008ci, Verma:2011yw, Demchuk:1997uz}, CCQM~\cite{Ivanov:2019nqd}, quark model (QM)~\cite{Isgur:1988gb, Melikhov:2000yu}, the large energy chiral quark model (LEChQM)~\cite{Palmer:2013yia}, relativistic harmonic oscillator potential model (RHOPM)~\cite{Wirbel:1985ji} and the HQETF~\cite{Wang:2002zba} are also used to study the TFFs of the corresponding process.

The rest of this paper is organized as follows. In Sec.~\ref{BFT}, we describe the calculation for the vector meson leading-twist longitudinal DAs $\phi_{2;V}^\parallel(x,\mu)$ within the BFTSRs in the scheme suggested in Ref.~\cite{Zhong:2021epq}. In Sec.~\ref{TFFs}, the TFFs and branching ratio of the semi-leptonic decays $D_{(s)}\to V$ within the LCSRs are presented. Numerical results are given in Sec.~\ref{biao2}. Sec.~\ref{biao3} is reserved for a summary.
\section{Theoretical Framework}\label{TF}

\subsection{$\xi$-moments of vector meson leading-twist longitudinal DA in the BFTSR}\label{BFT}

The light vector meson leading-twist longitudinal DAs $\phi^\parallel_{2;V}(x,\mu) (V = \rho, K^\ast, \phi)$ is defined as~\cite{Ball:2004rg},
\begin{align}
&\langle0|\bar{q}_1(z)\gamma_\mu q_2(-z)|V(q,\lambda)\rangle \nonumber\\
&\quad\quad = m_V f_V^\parallel \int_0^1 dx e^{i(2x-1)z\cdot q} q_\mu \frac{e^{(\lambda)}\cdot z}{q\cdot z} \phi_{2;V}^\parallel(x,\mu).
\label{eq:DA_definition}
\end{align}
In Eq.\eqref{eq:DA_definition}, $z^2 = 0$, $q, \lambda, m_V, f_V^\parallel$ and $e^{(\lambda)}$ are the momentum, helicity, mass, longitudinal decay constant and polarization vector of vector meson $V$, respectively. $q_1$ and $q_2$ indicate the vector meson valence quarks with $q_1 = d, q_2 = u$ for $\rho^+$, $q_1 = s, q_2 = d$ for $K^{\ast0}$, $q_1 = q_2 = s$ for $\phi$, respectively. $x$ is the momentum fraction of valance quark $q_1$ and $\mu$ stands for the scale. By expanding the left and right hand of Eq.\eqref{eq:DA_definition} respect to $z$, one can obtain the following matrix element definition of the $\xi$-moment $\langle\xi^{n}\rangle^{\|}_{2;V}$ of vector meson leading-twist longitudinal DA,
\begin{align}
\langle 0|\bar{q}_1(0)\slashed{z} (i z\cdot \tensor{D})^n q_2(0)|V(q,\lambda)\rangle = (z\cdot q)^{n+1}f^\parallel_V \langle\xi^n\rangle^\parallel_{2;V},
\label{eq:xi_moment}
\end{align}
with
\begin{align}
\langle\xi^n\rangle^\parallel_{2;V} = \int^1_0 dx (2x-1)^n \phi_{2;V}^\parallel(x,\mu).
\end{align}
In Eq.~\eqref{eq:xi_moment}, $\tensor{D}_\mu = \overrightarrow{D}_\mu - \overleftarrow{D}_\mu$ with the fundamental representation of the gauge covariant derivative $D_\mu = \partial_\mu - ig_s T^A \mathcal{A}_\mu^A (A = 1, \cdots, 8)$. It should be noted that the $\xi$-moment is scale $\mu$ dependent, i.e. $\langle\xi^n\rangle^\parallel_{2;V} \equiv \langle\xi^n\rangle^\parallel_{2;V}|_\mu$. For simplicity, we omit the subscript ``$\mu$'' in this paper. To derive the $\langle \xi^n \rangle_{2;V}^\parallel$, we then introduce the correlation function (correlator),
\begin{align}
\Pi_{2;V}(z,q) &= i\int d^{4}x e^{iq\cdot x} \langle0|T\{J_n(x),J_0^\dagger(0)\}|0\rangle \nonumber\\
&= (z\cdot q)^{n+2} I_{2;V}(q^2),
\label{eq:correlator}
\end{align}
with the interpolating currents
\begin{align}
J_n(x) &= \bar{q}_1(x) \slashed{z} (iz\cdot\tensor{D})^n q_2(x), \nonumber\\
J_0^\dagger(0) &= \bar{q}_2(0) \slashed{z} q_1(0).
\label{eq:currents}
\end{align}

Following the standard procedure of the SVZ QCD SRs, the correlator~\eqref{eq:correlator} can be treated by inserting a complete set of intermediate hadronic states in the physical region. With Eq.~\eqref{eq:xi_moment} and the quark-hadron daulity, the hadronic representation of correlator~\eqref{eq:correlator} can be obtained as
\begin{align}
{\rm Im} I_{2;V}^{\rm had}(s) &= \pi \delta(s - m_V^2) \langle\xi^n\rangle^\parallel_{2;V} \langle\xi^0\rangle^\parallel_{2;V}(f_V^\parallel)^2 \nonumber \\
&+ {\rm Im} I_{2;V}^{\rm pert}(s)\theta(s - s_V),
\end{align}
where $s_V$ indicates the continuum threshold parameter.
\begin{figure}[htp]
\includegraphics[width=0.35\textwidth]{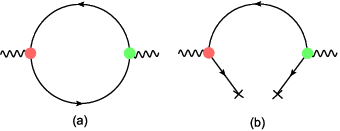}
\caption{Schematic Feynman diagrams for vector meson $\rho, K^\ast, \phi$ leading-twist longitudinal DAs $\xi$-moments. The left big dot and the right big dot indicate the vertex operators $\slashed z (iz\cdot\tensor{D})^n$ and $\slashed z$ from currents $J_n(x)$ and $J_0^\dagger(0)$, respectively. The cross symbol attached to the quark line stands for the local quark background field. The full Feynman diagrams can be found in Figs.~\ref{fig:FeynDiag_suba},\ref{fig:FeynDiag_subb}.}
\label{fig:FeynDiag}
\end{figure}
\begin{figure*}[htp]
\includegraphics[width=0.9\textwidth]{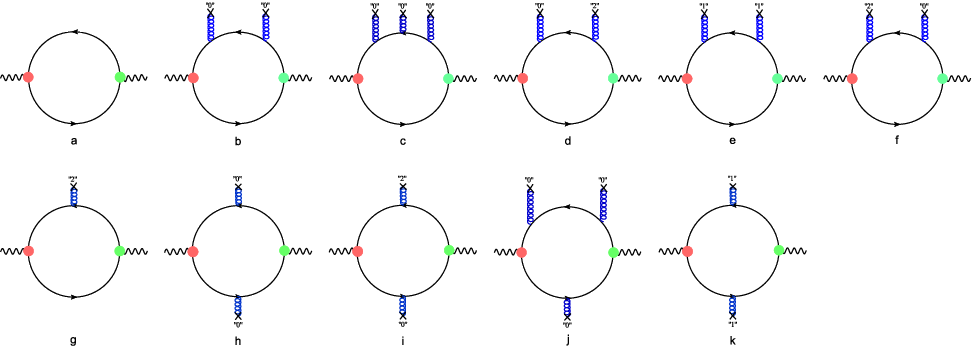}
\caption{Sub-diagrams of Fig.~\ref{fig:FeynDiag}(a), where the permutation diagrams have been omitted. The cross symbol attached to the gluon line stands for the local gluon background field strength tensor with ``$n$'' indicates $n$th order covariant derivative.}
\label{fig:FeynDiag_suba}
\end{figure*}
\begin{figure*}[htp]
\includegraphics[width=0.45\textwidth]{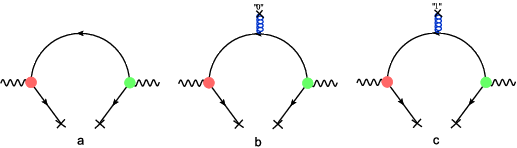}
\caption{Sub-diagrams of Fig.~\ref{fig:FeynDiag}(b).}
\label{fig:FeynDiag_subb}
\end{figure*}
On the other hand, we perform operator product expansion (OPE) on the correlator~\eqref{eq:correlator} in the deep Euclidean region. The corresponding calculation is performed in the framework of BFT. With the basic assumption and Feynman rules of BFT, the correlator~\eqref{eq:correlator} can be rewritten as
\begin{align}
\Pi_{2;V}(z,q) &= i\int d^4x e^{iq\cdot x} \nonumber\\
&\times \Big\{ - {\rm Tr} \langle0|S_F^{q_1}(0,x) \slashed z (iz\cdot\tensor{D})^n S_{F}^{q_2}(x,0) \slashed z|0\rangle  \nonumber \\
&+ {\rm Tr} \langle0| \bar q_1(x)q_1(0) \slashed z (iz\cdot\tensor{D})^n S_{F}^{q_2}(x,0) \slashed z |0\rangle \nonumber \\
&+ {\rm Tr} \langle0| S_{F}^{q_1}(0,x) \slashed z (iz\cdot\tensor{D})^n \bar q_2(0)q_2(x) \slashed z|0\rangle \Big\} \nonumber \\
&+\cdots,
\label{eq:OPE}
\end{align}
Where $\rm Tr$ indicates the trace of the color matrix and $\gamma$ matrix, ellipsis stands for ignored high-order corrections, $S_F^{q_1}(0,x)$ is the $q_1$-quark propagator from $x$ to $0$, $S_{F}^{q_2}(x,0)$ is the $q_2$-quark propagator from $0$ to $x$, $\slashed z (iz\cdot\tensor{D})^n$ and $\slashed z$ are the vertex operators from currents $J_n(x)$ and $J_0^\dagger(0)$ in Eq.~\eqref{eq:currents}, respectively. It should be noted that, in Eq.~\eqref{eq:OPE} the independent variable of the gluon field in the vertex operator $(iz\cdot\tensor{D})^n$ is $z$ instead of $x$, therefore we have $z\cdot\tensor{D} \equiv z\cdot\tensor{\partial}$ with fixed-point gauge $z^\mu \mathcal{A}_\mu^A = 0$~\cite{Shifman:1980ui}. The Feynman diagrams in Fig.~\ref{fig:FeynDiag}(a) and \ref{fig:FeynDiag}(b) correspond to the first and second term in Eq.~\eqref{eq:OPE}, while the permutation diagram of Fig.~\ref{fig:FeynDiag}(b) corresponds to the third term in Eq.~\eqref{eq:OPE}. The quark propagator in the background field up to dimension-six can be expressed as:
\begin{align}
S_F(x,0) &= S_F^0(x,0) + S_F^2(x,0) + S_F^3(x,0) + \sum_{i=1}^2 S_F^{4(i)}(x,0) \nonumber\\
&+ \sum_{i=1}^3 S_F^{5(i)}(x,0) + \sum_{i=1}^5 S_F^{6(i)}(x,0).
\label{eq:quarprop}
\end{align}
The specific expressions for each term in Eq.~\eqref{eq:quarprop} can refer to Ref.~\cite{Zhong:2014jla}. Then one can get the sub-diagrams of Fig.~\ref{fig:FeynDiag}(a) and \ref{fig:FeynDiag}(b), which are shown in Fig.~\ref{fig:FeynDiag_suba} and Fig.~\ref{fig:FeynDiag_subb} respectively. By matching the hadronic expression in the physical region and OPE in the deep Euclidean region of correlator~\eqref{eq:correlator} with the dispersion relation after Borel transformation,
\begin{align}
\frac{1}{\pi} \frac{1}{M^2} \int ds e^{-s/M^2} {\rm Im} I_{2;V}^{\rm had}(s)=\hat{\mathcal{B}}_{M^2} I_{2;V}^{\rm QCD}(q^2),
\label{eq:disprelt}
\end{align}
with the Borel parameter $M$ and the Borel transformation operator $\hat{\mathcal{B}}_{M^2}$, one can finally obtain,
\begin{widetext}
\begin{align}
\frac{\langle \xi^n \rangle^\parallel_{2;V} \langle \xi^0 \rangle^\parallel_{2;V} (f_V^\parallel)^2} {M^2 e^{m_V^2/M^2}} &=\frac{1}{\pi} \frac{1}{M^2} \int_{(m_1+m_2)^2}^{s_V} ds e^{-s/M^2} {\rm Im} I^{\rm pert}_{2;V}(s) + \frac{m_1 \langle \bar q_1 q_1 \rangle + (-1)^n m_2 \langle \bar q_2 q_2 \rangle} {(M^2)^2} - \frac{m_1m_2}{(M^2)^3} \Big[ m_2 \langle \bar{q}_1 q_1\rangle \nonumber\\
&+ (-1)^n m_1 \langle \bar q_2 q_2 \rangle \Big] +\frac{1+(-1)^n}{24\pi(n+1)} \frac{\langle\alpha_{s} G^{2}\rangle}{\mathrm{(M^2)}^2} - \frac{n}{3(M^2)^3} \Big[ m_1 \langle g_s \bar q_1 \sigma TGq_1\rangle + (-1)^{n} m_2 \langle g_s \bar q_2 \sigma TGq_2\rangle \Big] \nonumber\\
&+ \frac{(2n-1)m_1 m_2}{6(M^2)^4} \Big[ m_2 \langle g_s \bar q_1 \sigma TGq_1\rangle + (-1)^n m_1 \langle g_s \bar q_2 \sigma TGq_2\rangle \Big] - \frac{[1+(-1)^n] n\theta(n-1)}{192\pi^2} \frac{\langle g_s^3 fG^3\rangle} {(M^2)^3} \nonumber\\
&+ \frac{4(n+1)}{81(M^2)^3} \Big[ \langle g_s \bar q_1 q_1\rangle^2 + (-1)^{n} \langle g_s \bar q_2 q_2\rangle^2 \Big] - \frac{2(2n+1)}{81(M^2)^4} \Big[ m_2^2 \langle g_s \bar q_1 q_1\rangle^2 + m_1^2 (-1)^n \langle g_s\bar q_2 q_2\rangle^2 \Big] \nonumber\\
&+\frac{[1+(-1)^n] \langle g_s^2 \bar q q\rangle^2}{5832 \pi^2}\frac{2+\kappa^2}{(M^2)^3 } \Big\{ 516 \Big[ \psi(n+1)-\ln \frac{M^2}{\mu^2} + 2\gamma_E \Big] + \theta(n-2) \Big[ (306n \nonumber\\
&+ 267) \psi_3(n) - \frac{306}{n}(n+1)^2 \Big] + \theta(n-1) \Big[ 9\psi_4(n) - \frac{6}{n} (59 n^2 + 38) + 612n \Big( \psi(n+1) - \ln \frac{M^2}{\mu^2} \nonumber\\
&+ 2\gamma_E \Big) \big]-760 \Big\} + \hat I^{m^2}_{\langle G^2\rangle}(M^2) + \hat I^{m^2}_{\langle G^3 \rangle}(M^2) + \hat I^{m^2}_{\langle q^4\rangle}(M^2),
\label{eq:xiSR}
\end{align}
with
\begin{align}
{\rm Im} I^{\rm pert}_{2;V}(s)&= \frac{3}{8\pi}\frac{1}{(n+1)(n+3)} \Big\{ \Big( \frac{-m_1^2 + m_2^2 + \bar s \nu}{s} \Big)^{n+1} \Big[ 2(n+1) \frac{s^2 - (m_1^2 - m_2^2 - \bar s \nu)^2}{4s^2} + 1 \Big] \nonumber\\
&- \Big( \frac{-m_1^2+m_2^2-\bar{s}\nu}{s} \Big)^{n+1}\Big[ 2(n+1) \frac{s^2-(m_1^2 - m_2^2 + \bar s \nu)^2}{4s^2} + 1 \Big] \Big\},
\end{align}
\end{widetext}
where the expression of the function $\psi_{3(4)}$ can be seen in~\eqref{A4}, $\theta(n)$ is the step-function, $\psi(n)$ is the Spence-function, $m_{1(2)}$ is the $q_{1(2)}$ quark mass, $\gamma_E$ is Euler constant, $\bar s = s-(m_1-m_2)^2$, $\nu = \sqrt{1 - 4m_1m_2/\bar s}$, respectively. In specific calculation of OPE, the quark mass corrections $\mathcal{O}(m^2)$ proportional to $m_{1(2)}^2$ or $m_1 m_2$ have been considered. The third, sixth and ninth terms in Eq.~\eqref{eq:xiSR} are the mass corrections for the double-quark condensate $\langle\bar q_{1(2)}q_{1(2)}\rangle$, quark-gluon mixed condensate $\langle g_s\bar q_{1(2)}\sigma TGq_{1(2)}\rangle$ and four-quark condensate $\langle g_s\bar q_{1(2)} q_{1(2)}\rangle^2$, respectively. $\hat I^{m^2}_{\langle G^2\rangle}(M^2)$, $\hat I^{m^2}_{\langle G^3 \rangle}(M^2)$, and $\hat I^{m^2}_{\langle q^4\rangle}(M^2)$ in Eq.~\eqref{eq:xiSR} respectively indicates the mass corrections for the double-gluon condensate $\langle\alpha_sG^2\rangle$, triple-gluon condensate $\langle g_s^3fG^3\rangle$, and four-quark condensate $\langle g_s^2\bar qq\rangle^2$ with $q = u/d$ quark, and their specific expressions are listed in Appendix~\ref{appendix}.

By setting $n=0$ in Eq.~\eqref{eq:xiSR}, the sum rule for zeroth-order moment $\langle\xi^0\rangle_{2;V}^\parallel$ can be obtained, which is obviously dependent on the Borel parameter $M$. In order to obtain more accurate $n$th order $\xi$-moment $\langle\xi^n\rangle_{2;V}^\parallel$, we adopt the following formula as suggested in Ref.~\cite{Zhong:2021epq},
\begin{equation}
\langle\xi^n\rangle^\parallel_{2;V} = \frac{\langle\xi^n\rangle^\parallel_{2;V} \langle\xi^0\rangle^\parallel_{2;V} \Big|_{\textrm{from Eq.~\eqref{eq:xiSR}}}} {\sqrt{\left(\langle\xi^0\rangle^\parallel_{2;V}\right)^2} \Big|_{\textrm{from Eq.~\eqref{eq:xiSR} by } n = 0}}.
\label{eq:xiSR_new}
\end{equation}

Usually, one can obtain the Gegenbauer moments $a_{2;V}^{\parallel; n}$ through the following relationship between $a_{2;V}^{\parallel; n}$ and $\langle\xi^n\rangle_{2;V}^\parallel$,
\begin{align}
&a_{2;V}^{\parallel; 1} = \frac{5}{3}\langle\xi^1\rangle_{2;V}^{\|}, \nonumber\\
&a_{2;V}^{\parallel; 2} = \frac{35}{12}\langle\xi^2\rangle_{2;V}^{\|}-\frac{7}{12}, \nonumber\\
&\cdots. \label{eq:xia}
\end{align}
Traditionally, based on the obtained $a_{2;V}^{\parallel;n}$, the behaviors of light vector meson leading-twist longitudinal DAs can be constrained or obtained with appropriate phenomenological models. However, the high-order Gegenbauer moments $a_{2;V}^{\parallel; n} (n > 4)$ by Eq.~\eqref{eq:xia} are extremely unreliable due to the error of $\langle\xi^n\rangle_{2;V}^\parallel$ and its large coefficient. Therefore, traditional methods are difficult to obtain the precise behavior of $\phi_{2;V}^\parallel$. As suggested in Ref.~\cite{Zhong:2021epq}, we will use appropriate phenomenological model to fit as many $\langle\xi^n\rangle_{2;V}^\parallel$ calculated with Eq.~\eqref{eq:xiSR_new} as possible through the least squares method to determine the reliable and stable behaviors of $\phi_{2;V}^\parallel$. Specifically, we will calculate the first ten $\langle\xi^n\rangle_{2;V}^\parallel$ as the constraint conditions by referring to the analysis of the constraints of $\xi$-moments on pion leading-twist DA in Ref.~\cite{Zhong:2022lmn}, and select the following normalized power-law parametrization form (PLP model),
\begin{align}
\phi^\parallel_{2;V}(x, \mu) = \frac{\Gamma[\alpha^\parallel_{2;V} + \beta^\parallel_{2;V} + 2]}{\Gamma [\alpha^\parallel_{2;V} + 1] + \Gamma [\beta^\parallel_{2;V} + 1]} x^{\alpha^\parallel_{2;V}} (1-x)^{\beta^\parallel_{2;V}},
\label{eq:PLP_model}
\end{align}
to describe the behaviors of vector $\rho, K^\ast, \phi$ meson leading-twist longitudinal DAs. The model parameters in Eq.~\eqref{eq:PLP_model} are scale dependent, i.e. $\alpha^\parallel_{2;V} = \alpha^\parallel_{2;V}(\mu)$ and $\beta^\parallel_{2;V} = \beta^\parallel_{2;V}(\mu)$, and for simplicity, we do not explicitly show this. The optimal values of $\alpha^\parallel_{2;V}, \beta^\parallel_{2;V}$ can be obtained by minimizing the likelihood function,
\begin{equation}
\chi^2(\theta) = \sum_{i=1}^{10} \frac{(y_i - \mu(x_i,\theta))^2}{\sigma_i^2},
\label{eq:chi2}
\end{equation}
where the fitting parameter $\theta = (\alpha^\parallel_{2;V},\beta_{2;V}^\parallel)$, the mean function $\mu(x_i,\theta) (x_i\to n)$ indicates $\langle\xi^n\rangle_{2;V}^\parallel$ defined by combining Eqs.~\eqref{eq:xi_moment} and \eqref{eq:PLP_model}, the central values and corresponding errors of $\langle\xi^n\rangle_{2;V}^\parallel$ calculated with sum rule~\eqref{eq:xiSR_new} are regarded as the independent measurements $y_i$ and variance $\sigma_i$. The goodness of fit can be judged by the probability,
\begin{align}
P_{\chi^2_{\rm min}} = \int_{\chi^2_{\rm min}}^\infty f(y; n_d) dy,
\label{eq:Pchi2}
\end{align}
where $f(y; n_d) = 1/[\Gamma(n_d/2)2^{n_d/2}] y^{n_d/2-1}e^{-y/2}$ with the number of degrees of freedom $n_d$ is the probability density function of $\chi^2(\theta)$. It should be specifically pointed out that in the above fitting operation, the $\xi$-moments calculated from Eq.~\eqref{eq:xiSR_new} are treated as independent fitting samples, and this treatment is an approximate assumption. Their center values and errors come from the same input parameters, such as quark masses, condensates, continuum thresholds, etc. Therefore, there is a certain correlation between them. In this work, we do not introduce a full covariance matrix in the fitting process to characterize this correlation. This approximation will overestimate the number of effective independent constraints to a certain extent, so the goodness of fit probability only serves as a qualitative reference indicator for the fitting effect. Nevertheless, this is fully adequate for our primary objective of determining the optimal best-fit parameters.

\subsection{$D_{(s)}\to V$ TFFs within LCSRs}\label{TFFs}
The $D_{(s)}\to V$ hadron matrix element is usually parameterized as a vector TFFs $V(q^2)$ and three axial vector TFFs $A_{0,1,2}(q^2)$. Specifically,
\begin{align}
\langle V(p, \lambda) &|\bar{q}_1 \gamma_{\mu} (1-\gamma_5) c | D_{(s)}(p + q) \rangle =\! -\! ie_{\mu}^{*(\lambda)} (m_{D_{(s)}} \!+\! m_{V}) \nonumber \\
&\times A_{1}(q^{2})+ i (e^{*(\lambda)} \cdot q\,)\, \,\frac{(2p + q)_{\mu}}{m_{D_{(s)}} + m_{V}}\, A_{2}(q^{2}) + i q_{\mu} \nonumber \\
&\times(\,e^{*(\lambda\,)} \cdot q)\, \frac{2 m_{V}}{q^{2}}\,\left[\, A_{3}(q^{2}) - A_{0}(q^{2})\, \right] + \epsilon_{\mu \nu \alpha \beta} \nonumber \\
& \times e^{*(\lambda) \nu} q^{\alpha} p^{\beta} \frac{2 V(q^{2})}{m_{D_{(s)}} + m_{V}},
\label{eq:TFF_Definition}
\end{align}
where $m_{D_{(s)}}$ indicates the mass of the charmed $D_{(s)}$ meson, $p$ is the final state vector meson momentum, $q$ is the four momentum transfer, and $e^{\ast(\lambda)}$ stands for the polarization vector of the vector meson with $\lambda=(\perp,\|)$ corresponding to the transverse and longitudinal components, respectively.

In order to obtain the $D_{(s)}\to V(q,\lambda)$ TFFs, we introduce the following vacuum-to-vector meson state correlator,
\begin{align}
&\Pi_\mu(p,q) = i \int d^4x \, e^{i q \cdot x} \langle V(q,\lambda) | \mathrm{T}\big\{j_{\mu}, j^{L^\dagger}_D(0)\big\} | 0 \rangle,
\label{eq:TFF_correlator}
\end{align}
where the left-handed chiral currents $j_{\mu}(x)=\bar{q}_1(x)\gamma_\mu (1-\gamma_5)c(x)$, $j^{L^\dagger}_D(0)=i\bar{c}(0) (1-\gamma_5)q_2(0)$ are adopted. The advantage of this choice is that it can significantly suppress the contributions from chiral-odd DA (such as $\phi_{2;V}^\bot$, $\phi_{3;V}^{\|}$, $\psi_{3;V}^{\|}$...), while retaining the $\delta^0$-order dominant contributions from the chiral-even ones~\cite{Zhong:2021epq,Fu:2014uea}.

Following the standard LCSRs procedure, we deal with the correlator Eq.~(\ref{eq:TFF_correlator}) from two aspects. On the one hand, by inserting a complete set of hadronic states into Eq.~(\ref{eq:TFF_correlator}) in the time-like $q^2$ region, and further separating the pole term of the lowest pseudoscalar $D_{(s)}$ meson, one can obtain the hadronic representation of the correlator,
\begin{align}
\Pi_\mu^{\rm had}(p, q) &= e_{\mu}^{\ast(\lambda)} \Pi_{1}^{\rm had} + (e_{\mu}^{\ast(\lambda)} \cdot q)(2p + q)_{\mu} \Pi_{2}^{\rm had} \nonumber\\
&+ (e_{\mu}^{\ast(\lambda)} \cdot q) q_{\mu} \Pi_{3}^{\rm had} + i \epsilon_{\mu}^{\nu \alpha \beta} e_{\nu}^{\ast(\lambda)} q_{\alpha} p_{\beta} \Pi_{4}^{\rm had},
\label{eq:had}
\end{align}
where
\begin{align}
\Pi_i^{\rm had} &= \frac{m_{D_{(s)}}^2 f_{D_{(s)}} (m_{D_{(s)}} + m_{V})}{\hat m_j \big[m_{D_{(s)}}^2 - (p+q)^2\big]} \tilde{C}_i(q^2) \nonumber\\
&+ \int_{s_0}^{\infty} \frac{\rho_i^{\text{QCD}}}{s - (p+q)^2} ds + \mathrm{subtractions},
\end{align}
with
\begin{align}
&\tilde{C}_1 = A_1,\,\,\, \tilde{C}_2 = \frac{A_2}{(m_{D_{(s)}} + m_{V})^2},\nonumber\\
&\tilde{C}_3 = \frac{2m_{V}}{q^2(m_{D_{(s)}} + m_{V})} [A_3(q^2) - A_0(q^2)],\nonumber \\
&\tilde{C}_4 = \frac{2V(q^2)}{(m_{D_{(s)}} + m_{V})^2}.
\end{align}
In which, $f_{D_{(s)}}$ is the decay constant of the charmed $D_{(s)}$ meson, $s_0$ is the threshold parameter, $\hat m_1 = m_c$ corresponds to $D\to \rho, K^\ast$, $\hat m_2 = m_c+m_s$ corresponds to $D_s\to \phi$. In calculation, the Eq.~\eqref{eq:TFF_Definition}, the matrix element $\langle D_{(s)} | \bar{c}i \gamma_5 q_1 | 0 \rangle = m^2_{D_{(s)}}f_{D_{(s)}}/\hat m_j$ and the quark-hadron duality have been used.

On the other hand, in the deep Euclidean region, at $q^2 \ll m_c^2$ and $(p+q)^2 \ll m_c^2$ with the charm quark mass $m_c$, one can perform OPE on correlator~\eqref{eq:TFF_correlator}. The heavy charm quark fields can be contracted as propagator near the light-cone,
\begin{align}
&\langle 0 | T\{c(x)\bar{c}(0)\} | 0 \rangle = i \!\int \frac{d^4k}{(2\pi)^4} e^{-ik \cdot x} \bigg\{ \frac{\slashed k + m_c}{m_c^2 - k^2} -\! g_s \int_0^1\!\! dv \nonumber\\
&\times G^{\mu\nu}(\nu x) \left[ \frac{1}{2} \frac{\slashed k + m_c}{(m_c^2 - k^2)^2} \sigma_{\mu\nu} + \frac{\nu}{m_c^2 - k^2} x_\mu \gamma_\nu \right] \bigg\}.
\end{align}
It contains the contributions of two-particle and three-particle Fock states. Then the correlator~\eqref{eq:TFF_correlator} can be expressed as,
\begin{widetext}
\begin{align}
\Pi_{\mu}^{\text{QCD}}(p,q) = &m_c \int \frac{d^4x \, d^4k}{(2\pi)^4} e^{i(q-k)\cdot x} \Bigg\{\frac{1}{m_c^2 - k^2} \Bigg\{ 2k^\mu \langle V(p, \lambda) | \bar{s}(x) q_1(0) | 0 \rangle- 2i k^\nu \langle V(p, \lambda) | \bar{s}(x) \sigma_{\mu\nu} q_1(0) | 0 \rangle \nonumber\\
&- \epsilon_{\mu\nu\alpha\beta} k^\nu \langle V(p, \lambda) | \bar{s}(x) \sigma_{\alpha\beta} q_1(0) | 0 \rangle\Bigg\}- \int dv \bigg\{\frac{k^\nu}{(m_c^2 - k^2)^2} \Bigg[ -i \langle V(p, \lambda) | \bar{s}(x) g_s G_{\mu\nu}(vx) q_1(0) | 0 \rangle \nonumber\\
&- 2 \langle V(p, \lambda) | \bar{s}(x) \sigma_{\mu\alpha} g_s G^{\alpha\nu}(vx) q_1(0) | 0 \rangle+ 2i \langle V(p, \lambda) | \bar{s}(x) i g_s \tilde{G}_{\mu\nu}(vx) \gamma_5 q_1(0) | 0 \rangle
\nonumber\\
&+ \frac{2v x_{\alpha}}{m_c^2 - k^2}  - \langle V(p, \lambda) | \bar{s}(x) g_s G_{\mu\alpha}(vx) q_1(0) | 0 \rangle - i \langle V(p, \lambda) | \bar{s}(x) \sigma_{\mu\beta} g_s G^{\alpha\beta}(vx) q_1(0) | 0 \rangle\Bigg]\Bigg\} \Bigg\},
\end{align}
\end{widetext}
where $\tilde{G}_{\mu\nu}(vx)=\epsilon_{\mu\nu\alpha\beta}G^{\alpha\beta}(vx)/2$, $G^{\mu\nu}$ is the gluonic field strength, $g_s$ denotes the strong coupling constant. The nonlocal matrix elements can be expressed in terms of the meson DAs of various twists. The corresponding matrix elements can be found in Refs~\cite{Ball:2004rg, Fu:2014uea, Ball:2007zt}.

With the help of dispersion relationship and Borel transformation, the LCSRs expression of the required TFFs can be obtained as\footnote{In the massless lepton limit, only three form factors, $A_{1,2}(q^2)$, $V(q^2)$ contribute to the $D_{(s)}\to V$ semi-leptonic decay~\cite{Ball:1997rj}. Then we give the expressions for these three TFFs in our calculation.},\\
\begin{align}
A_1(q^2) &= \frac{2m_im_c m_V f_V^\|}{m_{D_{(s)}}^2 f_{D_{(s)}} (m_{D_{(s)}}\! +\! m_V)}\! \bigg\{\! \int_0^1 \frac{du}{u} e^{m_{D_{(s)}}^2-s(u) / M^2} \nonumber\\
&\quad \times \bigg[ \Theta(c(u, s_0)) \phi_{3,V}^{\perp}(u) \!-\! \frac{m_V^2}{u M^2} \widetilde{\Theta}(c(u, s_0)) C_V^\parallel(u) \bigg] \nonumber \\
&\quad - m_V^2 \!\int\! \mathcal{D}\underline{\alpha} \int dv~ e^{-s(X) / M^2} \frac{1}{X^2 M^2} \, \Theta(c(X, s_0))
\nonumber \\
&\quad \times \big[ \Phi_{3;V}^\parallel(\underline{\alpha}) + \widetilde{\Phi}_{3;V}^\parallel(\underline{\alpha}) \big] \bigg\},
\end{align}
\begin{align}
A_2(q^2) &= \frac{m_im_c m_V (m_{D_{(s)}}\! +\! m_V) f_V^\parallel}{m_{D_{(s)}}^2 f_{D_{(s)}}}~e^{m_{D_{(s)}}2 / M^2}~ \!\bigg\{2 \!\int_0^1 \frac{\mathrm{d}u}{u} \nonumber\\
&\quad \times e^{-s(u) / M^2} \bigg[ \frac{1}{u M^2} \widetilde{\Theta}(c(u, s_0)) A_V^\parallel(u) + \frac{m_V^2}{u M^4} \nonumber\\
&\quad \times \widetilde{\widetilde{\Theta}}(c(u, s_0)) C_V^\parallel(u) + \frac{m_b^2 m_V^2}{4 u^4 M^6} \widetilde{\widetilde{{\Theta}}}(c(u, s_0)) B_V^\parallel(u) \bigg] \nonumber\\
&\quad + m_V^2 \int \mathcal{D}\underline{\alpha} \int \mathrm{d}\nu \; e^{-s(X) / M^2} \; \frac{1}{X^3 M^4} \; \Theta(c(X, s_0)) \nonumber\\
&\quad \times \big[ \Phi_{3;V}(\widetilde{\underline{\alpha}}) + \widetilde{\Phi}_{3;V}^\parallel(\widetilde{\underline{\alpha}}) \big] \bigg\},
\end{align}
\begin{align}
V(q^2) &= \frac{m_im_c m_V (m_{D_{(s)}} \!+\! m_V) f^\parallel_V}{2m_{D_{(s)}}^2 f_{D_{(s)}}} e^{m_{D_{(s)}}^2 / M^2}\! \!\int_0^1\! \mathrm{d}u  e^{-s(u) / M^2} \nonumber\\
&\quad \times \frac{1}{u^2 M^2} \widetilde{\Theta}(c(u, s_0)) \psi_{3,V}^\perp(u),
\end{align}
where $s(\varrho)=[m_b^2-\bar \varrho(q^2-\varrho m_V^2)]/\varrho$ with $\bar \varrho = 1 - \varrho$ ($\varrho$ stands for $u$ or $X$), and $X=a_1 + v a_3$. $\Theta(c(\varrho,s_0))$ is the usual step function with $c(\varrho,s_0)=\varrho s_0 - m_b^2 + \bar \varrho q^2 - \varrho \bar \varrho m_V^2$.  $\mathcal{D}\underline{\alpha} = d\alpha_1 d\alpha_2 d\alpha_3 \delta(1-\alpha_1-\alpha_2-\alpha_3)$ with $\underline{\alpha}=\big\{\alpha_1,\alpha_2,\alpha_3\big\}$ corresponds to the momentum fractions carried by the quark, antiquark and gluon, respectively. $\widetilde\Theta (c(u,s_0))$, $\widetilde{\widetilde\Theta}(c(u,s_0))$ are defined via the integration as following,
\begin{widetext}
\begin{align}
&\int_{0}^{1}\frac{du}{u^{2}M^{2}}e^{-s(u)/M^{2}}\widetilde{\Theta}(c(u,s_{0}))f(u)=\int_{u_{0}}^{1}
\frac{du}{u^{2}M^{2}}e^{-s(u)/M^{2}}f(u)+\delta(c(u_{0},s_{0})),
\\[5pt]
&\int_{0}^{1}
\frac{du}{2u^{3}M^{4}}e^{-s(u)/M^{2}}
\widetilde{\widetilde{\Theta}}(c(u,s_{0}))f(u)=\int_{u_{0}}^{1}\frac{du}{2u^{3}M^{4}}e^{-s(u)/M^{2}}
f(u)+\Delta(c(u_{0},s_{0})),
\end{align}
\end{widetext}
where $\delta(c(u_0,s_0))=e^{-s_0/M^2}\frac{f(u_0)}{\mathcal{C}_0}$ and
\begin{align}
\Delta(c(u, s_0)) &= e^{-s_0/M^2} \bigg[\frac{1}{2u_0M^2} \frac{f(u_0)}{C_0} - \frac{u_0^2}{2C_0} \frac{d}{du}
\nonumber\\
&\times\left( \frac{f(u)}{uC} \right) \bigg|_{u=u_0} \bigg],
\end{align}
with $C_0 = m_b^2+u_0^2 m_V^2-q^2$ and $u_0\in [0,1]$ is the solution of $c(u_0, s_0)=0$. The simplified DAs $A_{V}^{\|}(u) = \int_{0}^{u} d\nu [\phi_{2;V}^{\|}(\nu) - \phi_{3;V}^{\perp}(\nu)]$, $B_V^\parallel(u) = \int_0^u d\nu\phi_{4;V}^\parallel(\nu)$ and $C_V^\parallel(\nu) = \int_0^u d\nu \int_0^{\nu} dw [\psi_{4;V}^\parallel(w)+ \phi_{2;V}^\parallel(w) - 2\phi_{3;V}^\parallel(w)]$ originate from the definition of the chiral-even
two-particle DAs~\cite{Ball:2004rg}. Here, $\phi_{3;V}^{\bot}$ and $\psi_{3;V}^{\bot}$ are twist-3 DAs, $\phi_{4;V}^{\|}$ and $\psi_{4;V}^{\|}$ are twist-4 DAs. The contributions from the three-particle twist-3 DAs $\widetilde{\Phi}_{3;V}^\parallel$, $\Phi_{3;V}^\parallel, \Phi_{3;V}(\widetilde{\underline{\alpha}})$, $\widetilde{\Phi}_{3;V}^\parallel(\widetilde{\underline{\alpha}})$ and the twist-4 DAs $\phi_{4;V}^{\|}$, $\psi_{4;V}^{\|}$ are negligible~\cite{Fu:2014cna}. For the $\phi_{3;V}^{\bot}$ and $\psi_{3;V}^{\bot}$ contributions in the $D_s^+\to\phi$ process, we follow the treatment suggested in Refs~\cite{Ball:2004rg}. For the $D\to \rho,K^*$ process, we use the Wandzura-Wilczek approximation~\cite{Ball:1997rj, Wandzura:1977qf}.
\begin{equation}
\begin{aligned}
 & \phi_{3;V}^{\perp}(x,\mu)=\frac{1}{2}\left[\int_{0}^{x}\mathrm{d}\nu\frac{\phi_{2;V}^{\parallel}(\nu,\mu)}
 {\bar{\nu}}+\int_{x}^{1}\mathrm{d}\nu\frac{\phi_{2;V}^{\parallel}(\nu,\mu)}{\nu}\right],
  \\
 & \psi_{3;V}^{\perp}(x,\mu)=2\left[\bar{x}\int_{0}^{x}\mathrm{d}\nu\frac{\phi_{2;V}^{\parallel}(\nu,\mu)}
 {\bar{\nu}}+x\int_{x}^{1}\mathrm{d}\nu\frac{\phi_{2;V}^{\parallel}(\nu,\mu)}{\nu}\right],
 \\
 & A_{V}^{\parallel}(x,\mu)=\frac{1}{2}\left[\bar{x}\int_{0}^{x}\mathrm{d}\nu\frac{\phi_{2;V}^{\parallel}
 (\nu,\mu)}{\bar{\nu}}+x\int_{x}^{1}\mathrm{d}\nu\frac{\phi_{2;V}^{\parallel}(\nu,\mu)}{\nu}\right],
\end{aligned}
\end{equation}

\begin{figure*}[htp]
\centering
\includegraphics[width=0.45\textwidth]{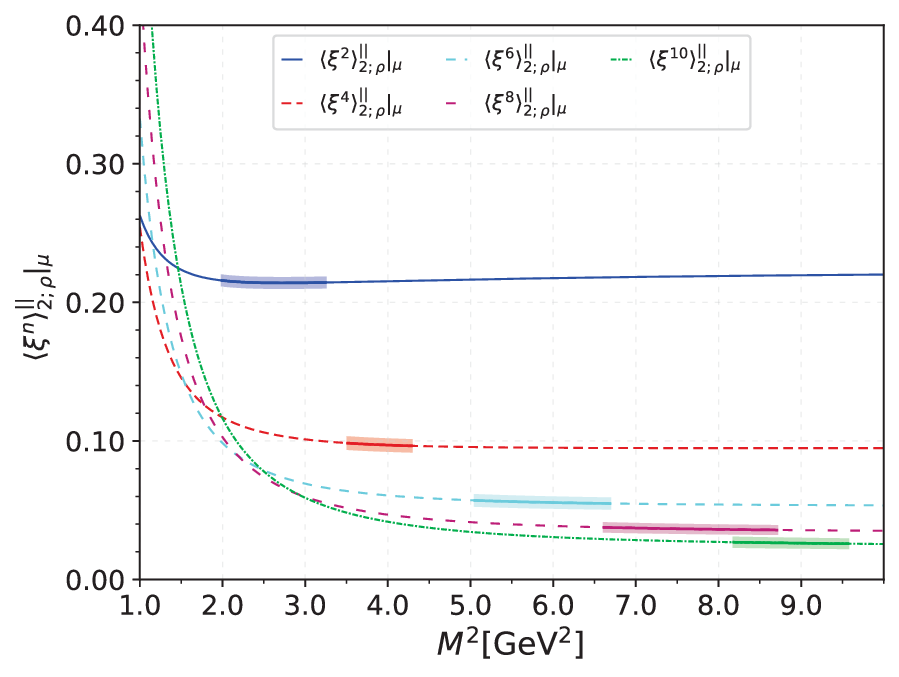}
\includegraphics[width=0.45\textwidth]{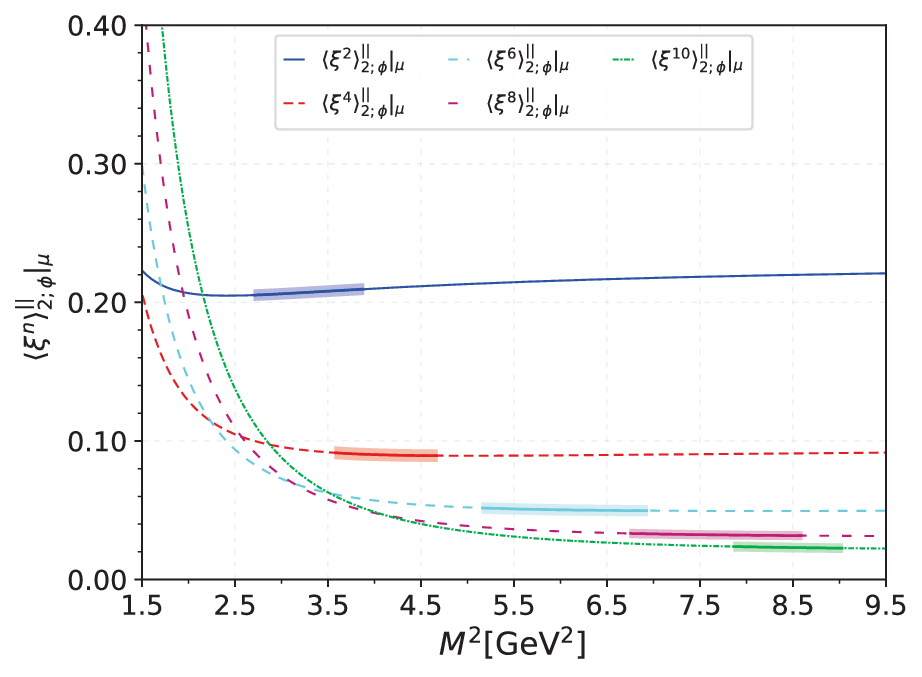}
\includegraphics[width=0.45\textwidth]{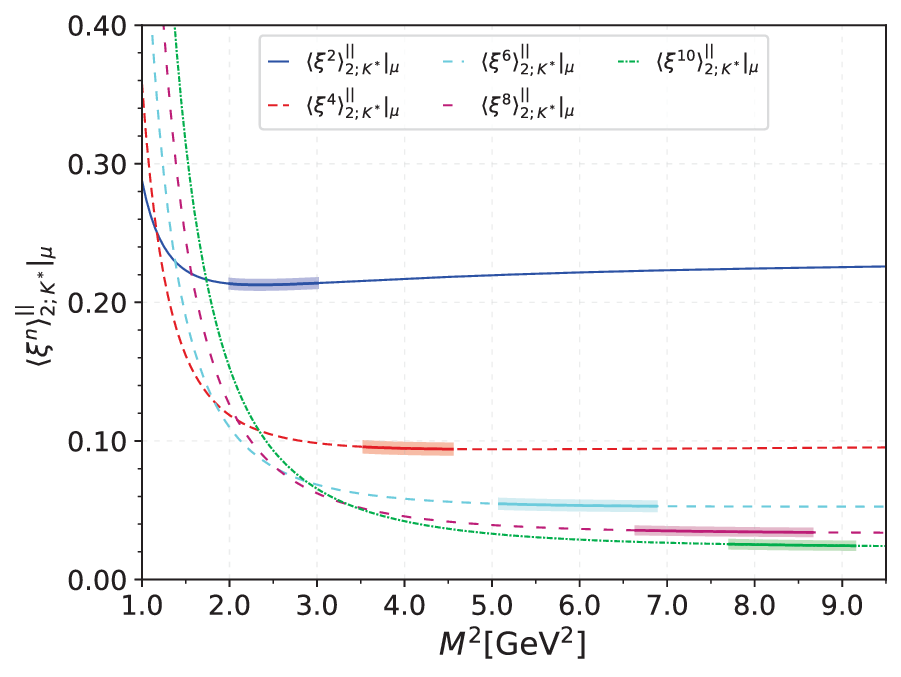}
\includegraphics[width=0.45\textwidth]{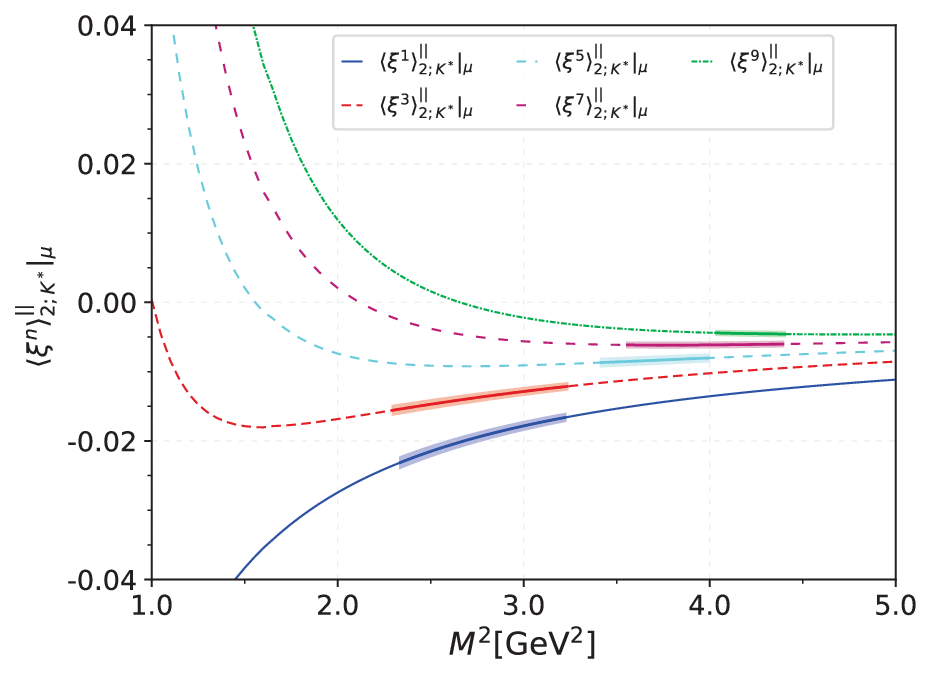}
\caption{$\xi$-moments $\langle\xi^n\rangle^\parallel_{2;V}~(n = 1,\cdots,10)$ with $V = \rho, K^\ast$ and $\phi$ versus the Borel parameter $M^2$. The shaded bands indicate the corresponding Borel windows.}
\label{fig:xi}
\end{figure*}

Based on this, we can further calculate the physical observables, e.g., the decay widths and branching fractions, for the relevant semi-leptonic decay processes. According to the polarization structure of a vector meson, there are three polarization states, i.e., two transverse polarization states (corresponding to helicities $\pm1$) and one longitudinal polarization state (helicity $0$). The expression for the differential decay width of the longitudinally polarized meson is given as follows~\cite{Fu:2014cna},
\begin{align}
\frac{d\Gamma_L}{dq^2} &= \frac{G_F^2 |V_{cq_i}|^2}{192\pi^2 m_{D_{(s)}}^3} \, \lambda^{1/2}(m_{D_{(s)}}^2, m_{V}^2, q^2) \, q^2\Bigg| \frac{1}{2m_{V}\sqrt{q^2}} \nonumber\\
&\times \bigg[ (m_{D_{(s)}}^2 - m_{V}^2 - q^2)(m_{D_{(s)}}^2 + m_{V}) A_1(q^2) \nonumber\\
&- \frac{\lambda(m_D^2, m_{V}^2, q^2)}{m_{D_{(s)}}^2 + m_{V}} A_2(q^2) \bigg] \Bigg|^2,
\label{eq:GamL}
\end{align}
where $\lambda(m_{D_{(s)}}^2, m_{V}^2, q^2)\!=\!(m_{D_(s)}^2+m_V^2-q^2)^2-4m_{D_{(s)}}^2 m_V^2$. The transverse differential decay width reads~\cite{Fu:2014cna},
\begin{align}
\frac{d\Gamma_T^{\pm}}{dq^2} &= \frac{G_F^2 |V_{cq_i}|^2}{192\pi^2 m_{D_{(s)}}^3}\lambda^{1/2}(m_{D_{(s)}}^2, m_V^2, q^2) \, q^2\bigg|(m_{D_{(s)}} + m_V) \nonumber\\
&\times A_1(q^2)\mp \frac{\lambda^{1/2}(m_{D_{(s)}}^2, m_V^2, q^2)}{m_{D_{(s)}} + m_V} V(q^2)
\bigg|^2.
\label{eq:GamT}
\end{align}
In the decay width expression provided above, $C_F$ is the Fermi constant. Depending on the specific decay process, $V_{cq_i}$ corresponds to the Cabibbo-Kobayashi-Maskawa (CKM) matrix element $V_{cd}$ or $V_{cs}$. The symbols ``$-$'' and ``$+$'' denote the right-handed and left-handed states, respectively. The total transverse decay width is given by $\Gamma_T=\Gamma^+_T+\Gamma^-_T$, and the total decay width is $\Gamma=\Gamma_L+\Gamma_T$. Finally, one can obtain the branching fraction by
\begin{align}
\mathcal{B}(D_{(s)} \to V \ell^+ \nu_\ell) = \tau_{D_{(s)}} \int_{m_\ell^2}^{(m_{D_{(s)}} - m_V)^2} \frac{d\Gamma}{dq^2},
\label{eq:Dec_Fra}
\end{align}
where, $m_{\ell}$ is the lepton mass with $\ell = e, \mu$, and $\tau_{D_{(s)}}$ is the lifetime of the $D_{(s)}$ meson.

\section{Numerical Analysis}\label{biao2}
\subsection{Input parameters}
\begin{table*}
\renewcommand{\arraystretch}{1.2}
\footnotesize
\caption{Our predictions for the first five nonzero $\xi$-moments $\langle\xi^n\rangle^\parallel_{2;V}(n=2, \cdots,10)$ and the second Gegenbauer moment $a^{2;\parallel}_{2;V}$  with $V = \rho, \phi$, compared to other theoretical predictions.}
\begin{tabular}{ @{}l
    @{\hspace{10pt}}c
    @{\hspace{15pt}}c
    @{\hspace{15pt}}c
    @{\hspace{15pt}}c
    @{\hspace{15pt}}c
    @{\hspace{15pt}}c
    @{\hspace{15pt}}c
    @{}}
\hline\hline
&$\mu[\rm{GeV}]$  &$\langle\xi^2\rangle^{\|}_{2;\rho}$  &$\langle\xi^4\rangle^{\|}_{2;\rho}$   &$\langle\xi^6\rangle^{\|}_{2;\rho}$    &$\langle\xi^8\rangle^{\|}_{2;\rho}$ &$\langle\xi^{10}\rangle^{\|}_{2;\rho}$     &$a^{2;\|}_{2;\rho}$\\\hline
This work  &1         & $0.225^{+0.013}_{-0.012}$      &$0.106^{+0.008}_{-0.008}$   & $0.063^{+0.006}_{-0.006}$  &$0.042^{+0.005}_{-0.005}$ &$0.032^{+0.004}_{-0.004}$  &$0.074^{+0.039}_{-0.036}$\\[1pt]
This work  &1.4       & $0.222^{+0.011}_{-0.011}$      &$0.102^{+0.007}_{-0.006}$   & $0.060^{+0.005}_{-0.005}$  &$0.040^{+0.004}_{-0.004}$ &$0.030^{+0.003}_{-0.003}$  &$0.063^{+0.033}_{-0.031}$\\[1pt]
This work  &2         & $0.219^{+0.010}_{-0.009}$      &$0.100^{+0.005}_{-0.005}$   & $0.058^{+0.004}_{-0.004}$  &$0.039^{+0.003}_{-0.003}$ &$0.029^{+0.003}_{-0.002}$  &$0.056^{+0.029}_{-0.027}$\\[1pt]
LQCD~\cite{Arthur:2010xf}    &2       &$0.25(2)(2)$         &$-$               &$-$               &$-$       &$-$     &$0.20(6)$\\[1pt]
LQCD~\cite{Boyle:2008nj}     &2       &$0.237(36)(12)$      &$-$               &$-$               &$-$       &$-$     &$0.108(105)(35)$ \\[1pt]
LQCD~\cite{Braun:2016wnx}    &2       &$-$                  &$-$               &$-$               &$-$       &$-$      &$0.132(27)$ \\[1pt]
BSWF~ \cite{Serna:2022yfp}   &2       &$0.263$             &$0.136$           &$0.090$           &$0.062$   &$-$       &$0.003\pm0.038$\\[1pt]
DSE  ~\cite{Gao:2014bca}     &2       &$0.23$              &$0.11$            &$0.066$           &$0.045$   &$0.033$    &$-0.0017$\\[1pt]
LFQM ~\cite{Choi:2007yu}     &1       &$0.21[0.19]$        &$0.09[0.08]$      &$0.05[0.04]$      &$-$       &$-$        &$0.02[-0.02]$\\[1pt]
LFQM ~\cite{Choi:2013mda}     &1       &$0.193(0.200)$      &$0.078[0.086]$    &$-$              &$-$       &$-$         &$-$\\[1pt]
LFQM ~\cite{Arifi:2025olq}     &1       &$0.193$             &$0.078$          &$0.040$          &$-$       &$-$          &$-0.020$\\[1pt]
QCD SRs~\cite{Pimikov:2013usa}&1       &$0.216(21)$         &$0.089(9)$        &$0.048(5)$   &$0.030(3)$&$0.022(2)$    &$0.047\pm0.058$\\[1pt]
QCD SRs~\cite{Bakulev:1998pf} &1       &$0.227(7)$     &$0.095(5)$     &$0.051(4)$    &$0.030(2)$  &$0.020(5)$    &$-$\\[1pt]
QCD SRs ~\cite{Ball:1996tb}   &1       &$-$               &$-$              &$-$               &$-$       &$-$             &$0.18\pm0.10$\\[1pt]
QCD SRs~\cite{Stefanis:2015qha} &1       &$0.206(8)$        &$0.087(6)$        &$-$               &$-$       &$-$              &$0.017(24)$\\[1pt]
LCQM ~\cite{Ji:1992yf}       &1       &$0.198$             &$0.083$           &$0.044$           &$-$       &$-$      &$-$    \\[1pt]
AdS/QCD ~\cite{Forshaw:2012im} &1     &$0.228$             &$-$               &$-$               &$-$       &$-$     &$-$   \\[1pt]
LFH ~\cite{Gurjar:2024wpq}  &1         &$0.20$             &$0.087$            &$0.048$           &$0.031$   &$0.022$     &$-$  \\[1pt]
\hline\hline
&$\mu[\rm{GeV}]$  &$\langle\xi^2\rangle^{\|}_{2;\phi}$  &$\langle\xi^4\rangle^{\|}_{2;\phi}$   &$\langle\xi^6\rangle^{\|}_{2;\phi}$    &$\langle\xi^8\rangle^{\|}_{2;\phi}$ &$\langle\xi^{10}\rangle^{\|}_{2;\phi}$     &$a^{2;\|}_{2;\phi}$\\\hline
This work$$  &1         & $0.209^{+0.020}_{-0.020}$     &$0.092^{+0.025}_{-0.025}$    &$0.052^{+0.017}_{-0.017}$    &$0.0335^{+0.0140}_{-0.0140}$     &$0.0250^{+0.0135}_{-0.0134}$    &$0.027^{+0.058}_{-0.058}$\\[1pt]
This work$$  &1.5       & $0.208^{+0.017}_{-0.017}$     &$0.091^{+0.019}_{-0.019}$    &$0.051^{+0.013}_{-0.013}$    &$0.0329^{+0.0100}_{-0.0100}$     &$0.0239^{+0.0094}_{-0.0093}$    &$0.023^{+0.048}_{-0.049}$\\[1pt]
This work$$  &2         & $0.207^{+0.015}_{-0.015}$     &$0.090^{+0.017}_{-0.017}$    &$0.050^{+0.011}_{-0.011}$    &$0.0327^{+0.0083}_{-0.0082}$     &$0.0235^{+0.076}_{-0.0076}$     &$0.011^{+0.047}_{-0.046}$\\[1pt]
LQCD~\cite{Arthur:2010xf}    &2       &$0.24(1)(1)$        &$-$               &$-$               &$-$       &$-$     &$-$\\[1pt]
LQCD~\cite{Boyle:2008nj}    &2       &$0.246(10)(12)$      &$-$               &$-$               &$-$       &$-$     &$-$ \\[1pt]
LQCD~\cite{Hua:2020gnw}    &2       &$0.212$              &$0.097$            &$0.057$          &$0.039$       &$-$     &$-$ \\[1pt]
QCD SRs ~\cite{Ball:2004rg}   &1       &$-$                 &$-$                &$-$              &$-$      &$-$    &$0.06^{+0.09}_{-0.07}$    \\[1pt]
QCD SRs ~\cite{Ball:2007zt}  &2       &$0.245$               &$0.115$           &$0.068$          &$0.045$      &$-$    &$0.18(8)$    \\[1pt]
QCD SRs ~\cite{Ball:1998sk}   &1       &$-$                 &$-$                &$-$              &$-$      &$-$    &$0\pm0.1$    \\[1pt]
 BSWF~ \cite{Serna:2022yfp}   &2       &$0.186$             &$0.077$           &$0.042$           &$0.026$   &$-$       &$-0.372\pm0.010$\\[1pt]
DSE  ~\cite{Gao:2014bca}     &2       &$0.233$              &$0.111$            &$0.067$           &$0.046$   &$-$    &$-$\\[1pt]
\hline\hline
\end{tabular}
\label{tab:xi_value1}
\end{table*}

\begin{table*}[htp]
\renewcommand{\arraystretch}{1.2}
\footnotesize
\caption{Our predictions for the first ten $\xi$-moments $\langle\xi^n\rangle^{\|}_{2;K^\ast}(n=1,...10)$ and the Gegenbauer moments $a^{2;\parallel}_{1;K^\ast}, a^{2;\parallel}_{2;K^\ast}$, compared to other theoretical predictions.}
\begin{tabular}{ @{}l
    @{\hspace{10pt}}c
    @{\hspace{15pt}}c
    @{\hspace{15pt}}c
    @{\hspace{15pt}}c
    @{\hspace{15pt}}c
    @{\hspace{15pt}}c
    @{\hspace{15pt}}c
    @{}}
\hline\hline
&$\mu[\rm{GeV}]$  &$\langle\xi^2\rangle^{\|}_{2;K^*}$  &$\langle\xi^4\rangle^{\|}_{2;K^*}$   &$\langle\xi^6\rangle^{\|}_{2;K^*}$    &$\langle\xi^8\rangle^{\|}_{2;K^*}$ &$\langle\xi^{10}\rangle^{\|}_{2;K^*}$     &$a^{2;\|}_{2;K^*}$\\
\hline
This work$$  &1       & $0.217^{+0.007}_{-0.007}$   &$0.098^{+0.010}_{-0.009}$   &$0.056^{+0.013}_{-0.012}$   &$0.037^{+0.018}_{-0.017}$ &$0.027^{+0.026}_{-0.025}$    &$0.050^{+0.020}_{-0.019}$ \\[1pt]
This work$$  &1.4       & $0.215^{+0.006}_{-0.006}$   &$0.096^{+0.008}_{-0.008}$   &$0.055^{+0.011}_{-0.010}$   &$0.036^{+0.014}_{-0.014}$ &$0.026^{+0.020}_{-0.019}$    &$0.043^{+0.017}_{-0.016}$ \\[1pt]
This work$$  &2       & $0.213^{+0.005}_{-0.005}$   &$0.095^{+0.007}_{-0.007}$   &$0.054^{+0.009}_{-0.008}$   &$0.035^{+0.012}_{-0.011}$ &$0.025^{+0.016}_{-0.015}$    &$0.038^{+0.015}_{-0.014}$ \\[1pt]
LQCD~\cite{Arthur:2010xf}    &2       &$0.25(1)(2)$        &$-$               &$-$               &$-$       &$-$     &$-$\\[1pt]
LQCD~\cite{Hua:2020gnw}    &2       &$0.205(3)$             &$-$           &$-$           &$-$   &$-$     &$-$ \\[1pt]
DSE ~ \cite{Xu:2025hjf}   &2       &$0.220$             &$-$               &$-$               &$-$      &$-$       &$-$\\[1pt]
  BSWF~\cite{Serna:2022yfp}     &2       &$0.272$              &$0.146$            &$0.097$     &$0.072$    &$-$    &$-0.191\pm0.048$\\[1pt]
LFQM ~\cite{Choi:2007yu}     &1       &$0.19[0.18]$        &$0.08[0.07]$      &$0.04[0.03]$      &$-$       &$-$        &$-0.03[-0.07]$\\[1pt]
LFQM ~\cite{Choi:2013mda}     &1       &$0.177$             &$0.067$          &$-$              &$-$       &$-$         &$-$\\[1pt]
LCQM ~\cite{Ji:1992yf}       &1       &$0.20$              &$0.08$            &$0.04$           &$-$       &$-$      &$-$    \\[1pt]
QCD SRs~\cite{Lin:2025cmn}    &2       &$-$                  &$-$               &$-$               &$-$       &$-$     &$0.16(9)$ \\[1pt]
QCD SRs ~\cite{Ball:2004rg}   &1      &$0.22^{+0.03}_{-0.02}$    &$0.10\pm0.02$&$-$               &$-$       &$-$      &$0.07^{+0.09}_{-0.07}$    \\[1pt]
QCD SRs ~\cite{Ball:2007zt}    &2       &$0.227$             &$0.104$           &$0.060$           &$0.039$    &$-$      &$0.11(9)$    \\[1pt]
QCD SRs ~\cite{Ball:2003sc}    &1       &$-$                &$-$                &$-$               &$-$        &$-$      &$0.09\pm0.05$    \\[1pt]
\hline\hline
&$\mu[\rm{GeV}]$  &$\langle\xi^1\rangle^{\|}_{2;K^*}$  &$\langle\xi^3\rangle^{\|}_{2;K^*}$   &$\langle\xi^5\rangle^{\|}_{2;K^*}$    &$\langle\xi^7\rangle^{\|}_{2;K^*}$ &$\langle\xi^{9}\rangle^{\|}_{2;K^*}$     &$a^{1;\|}_{2;K^*}$\\\hline
This work$$  &1         &$-0.0228^{+0.0042}_{-0.0040}$    &$-0.0169^{+0.0049}_{-0.0044}$ &$-0.0108^{+0.0054}_{-0.0046}$     &$-0.0087^{+0.0067}_{-0.0053}$       &$-0.0071^{+0.0084}_{-0.0065}$     &$-0.038^{+0.007}_{-0.007}$\\[1pt]
This work$$  &1.4       &$-0.0206^{+0.0038}_{-0.0036}$     &$-0.0146^{+0.0042}_{-0.0037}$   &$-0.0092^{+0.0045}_{-0.0039}$     &$-0.0071^{+0.0054}_{-0.0043}$        &$-0.0056^{+0.0066}_{-0.0051}$     &$-0.034^{+0.006}_{-0.006}$\\[1pt]
This work$$  &2          &$-0.0190^{+0.0035}_{-0.0033}$     &$-0.0131^{+0.0037}_{-0.0033}$   &$-0.0083^{+0.0039}_{-0.0033}$      &$-0.0061^{+0.0045}_{-0.0036}$        &$-0.0048^{+0.0054}_{-0.0042}$     &$-0.032^{+0.006}_{-0.006}$\\[1pt]
LQCD~\cite{Arthur:2010xf}    &2       &$0.037(1)(2)$        &$-$               &$-$               &$-$       &$-$     &$-$\\[1pt]
LQCD~\cite{Hua:2020gnw}    &2       &$0.003(4) $           &$-$               &$-$               &$-$       &$-$     &$-$ \\[1pt]
DSE ~ \cite{Xu:2025hjf}   &2       &$0.023$             &$-$               &$-$               &$-$       &$-$       &$-$\\[1pt]
BSWF~\cite{Serna:2022yfp}     &2       &$-$                  &$-$               &$-$               &$-$       &$-$     &$0.041\pm0.027$ \\[1pt]
LFQM ~\cite{Choi:2007yu}     &1       &$0.07[0.09]$        &$0.03[0.04]$      &$0.02[0.02]$      &$-$       &$-$        &$0.11[0.14]$\\[1pt]
LFQM ~\cite{Choi:2013mda}     &1       &$0.085$             &$0.040$           &$-$              &$-$       &$-$         &$-$\\[1pt]
LCQM ~\cite{Ji:1992yf}       &1       &$0.046$             &$0.025$           &$0.015$           &$-$       &$-$      &$-$    \\[1pt]
QCD SRs~\cite{Lin:2025cmn}    &2       &$-$                  &$-$               &$-$               &$-$       &$-$     &$0.06(4)$ \\[1pt]
QCD SRs~\cite{Ball:2004rg}        &1       &$0.06\pm0.04$       &$-$                &$-$              &$-$       &$-$      &$0.10\pm0.07$    \\[1pt]
QCD SRs ~\cite{Ball:2007zt}        &2       &$-$               &$-$                &$-$              &$-$       &$-$      &$0.03(2)$    \\[1pt]
QCD SRs ~\cite{Ball:2003sc}        &1       &$-$               &$-$                &$-$              &$-$       &$-$      &$-0.40\pm0.20$    \\[1pt]
\hline\hline
\end{tabular}
\label{tab:xi_value2}
\end{table*}

To obtain the numerical results for the leading-twist longitudinal DA of the $\rho$, $K^*$, and $\phi$ mesons. We take the decay constants $f_{\rho}=205\pm9~\rm{MeV}$, $f_{K^*}=217\pm5~\rm{MeV}$ and $f_{\phi}=231\pm4~\rm{MeV}$~\cite{Ball:2004rg}, the meson mass $m_{\rho}=775.26\pm0.23~\rm{MeV}$, $m_{K^\ast}=891.67\pm0.26~\rm{MeV}$ and $m_{\phi}=1019.460\pm0.016~\rm{MeV}$, the quark mass $m_u (2~{\rm GeV}) = 2.16 \pm 0.07~\rm{MeV}$,$m_d  (2~{\rm GeV}) = 4.70 \pm 0.07~\rm{MeV}$, $m_s  (2~{\rm GeV}) = 93.5 \pm 0.8~\rm{MeV}$ and $\bar m_c(\bar m_c)=1.2730\pm0.0046~\rm{GeV}$ from PDG \cite{ParticleDataGroup:2024cfk}. The values of the non-perturbative vacuum condensates at $\mu = 2~\rm GeV$ can be obtained from Ref.~\cite{Zhong:2021epq},
\begin{align}
&\langle \bar{q}q\rangle=(-2.417^{+0.227}_{-0.114})\times10^{-2}~\rm{GeV^2},
\nonumber\\
&\langle g_s \bar{q}\sigma TGq\rangle=(-1.934^{+0.188}_{-0.103})\times10^{-2}~\rm{GeV^5},
\nonumber\\
&\langle g_s \bar{s}\sigma TGs\rangle=\kappa\langle g_s     \bar{q}\sigma TGq\rangle,
\nonumber\\
&\langle g_s \bar{q} q\rangle^2=(2.082^{+0.734}_{-0.697})\times10^{-3}~\rm{GeV^6},
\nonumber\\
&\langle g_s^2 \bar{q} q\rangle^2=(7.420^{+2.614}_{-2.483})\times10^{-3}~\rm{GeV^6},
\nonumber\\
&\langle \alpha_sG^2\rangle=0.038\pm0.011~\rm{GeV^4},
\nonumber\\
&\langle g_s^3fG^2\rangle\simeq0.045~\rm{GeV^6},
\nonumber\\
&\langle \bar{s}s\rangle=\kappa\langle\bar{q}q\rangle,
\end{align}
with the ratio $\kappa=0.74\pm0.03$ \cite{Narison:2014wqa}. The scale-dependence of those inputs are exhibited as follows~\cite{Zhong:2021epq},
\begin{align}
&{m}_{q(s)}(\mu)={m}_{q(s)}(\mu_{0})\left[\frac{\alpha_{s}(\mu_{0})}{\alpha_{s}(\mu)}\right]^{-4/\beta_{0}}, \nonumber\\
&\bar{m}_{c}(\mu)=\bar{m}_{c}(\bar{m}_{c})\left[\frac{\alpha_{s}(\mu_{0})}{\alpha_{s}(\mu)}\right]^{-4/\beta_{0}},
\nonumber\\
&\langle \bar{q}q\rangle(\mu)=\langle\bar{q}q\rangle(\mu_{0})\left[\frac{\alpha_{s} (\mu_{0})}{\alpha_{s} (\mu)}\right]^{4/\beta_{0}}, \nonumber\\
&\langle g_{s}\bar{q}q\rangle^{2}(\mu)=\langle g_{s}\bar{q}q\rangle^{2}\left[\frac{\alpha_{s} (\mu_{0})}{\alpha_{s} (\mu)}\right]^{-2/(3 \beta_{0})}, \nonumber\\
&\langle g_{s}\bar{q}\sigma TGq\rangle(\mu)=\langle g_{s}\bar{q}\sigma TGq\rangle(\mu_{0})\left[\frac {\alpha_{s}(\mu_{0})}{\alpha_{s}(\mu)}\right]^{-2/(3\beta_{0})}, \nonumber\\
&\langle g_{s}^{2}\bar{q}q\rangle^{2}(\mu)=\langle g_{s}^{2}\bar{q}q\rangle^{2}(\mu_{0}), \nonumber\\
&\langle\alpha_{s}G^{2}\rangle(\mu)=\langle\alpha_{s}G^{2}\rangle(\mu_{0}), \nonumber\\
&\langle g_{s}^{3}fG^{3}\rangle(\mu)=\langle g_{s}^{3}fG^{3}\rangle(\mu_{0}),
\end{align}
where $\beta_0 = (33-2n_f)/3$ with the number of quark flavors $n_f$. By requiring that there is a reasonable Borel window to normalize $\langle\xi^0\rangle_{2;V}^\parallel$ with Eq.~\eqref{eq:xiSR} by taking $n = 0$, the continuous threshold parameters can be obtained as $s_\rho = 2.1\rm~GeV^2$, $s_{K^\ast} = 2.6\rm ~GeV^2$, $s_\phi = 2.9\rm ~GeV^2$.

\subsection{$\xi$-moments of vector meson leading-twist longitudinal DAs}
Then, one can obtain the curves of $\xi$-moments $\langle\xi^n\rangle^\parallel_{2;V} (V = \rho, K^\ast, \phi)$ versus Borel parameter $M^2$ (see Fig.~\ref{fig:xi}) with Eq.~\eqref{eq:xiSR_new}. In order to obtain the values of $\langle\xi^n\rangle^\parallel_{2;V}$, the corresponding Borel windows should be determined. In principle, the contributions of continuous state and six-dimensional condensation are required to be as small as possible, while the $\xi$-moments need to remain stable in the Borel windows. Specifically, the continuum contributions are required to be below $45\%$ for $\langle\xi^n\rangle^\parallel_{2;V} (n\le6)$ and $50\%$ for $\langle\xi^n\rangle^\parallel_{2;V} (n\ge8)$ while the dimension-six condensate contributions are required to be less than $5\%$ for $\langle\xi^n\rangle^\parallel_{2;V} (n\le 6)$ and $8\%$ for $\langle\xi^n\rangle^\parallel_{2;V} (n\ge8)$. The Borel windows of $\langle\xi^n\rangle^\parallel_{2;V}$ are shown in Fig.~\ref{fig:xi}, where all the inputs are taken to be their central values. Based on this, we obtain the values of the $\xi$-moments $\langle\xi^n\rangle^\parallel_{2;V} (V = \rho, K^\ast, \phi)$ up to the tenth order as well as the Gegenbauer moments $a_{2;K^\ast}^{1;\parallel}$ and $a_{2;V}^{2;\parallel}$ at scales $\mu = 1, 1.4, 2~{\rm GeV}$, and which are exhibited in Table~\ref{tab:xi_value1} and Table~\ref{tab:xi_value2}, respectively. The corresponding errors come from the squared average for the Borel parameter and other inputs such as meson decay constants, meson masses, quark masses and vacuum condensates, etc. As a comparison, those moments by various methods such as LQCD~\cite{Arthur:2010xf, Boyle:2008nj, Braun:2016wnx, Hua:2020gnw}, BSWF~ \cite{Serna:2022yfp}, QCD SRs~\cite{Ball:1998sk, Ball:2004rg, Pimikov:2013usa, Bakulev:1998pf, Ball:1996tb, Stefanis:2015qha, Ball:2007zt, Ball:2003sc,Lin:2025cmn}, DSE~\cite{Gao:2014bca, Xu:2025hjf}, LFQM~\cite{Choi:2007yu, Choi:2013mda, Arifi:2025olq}, LCQM~\cite{Ji:1992yf}, AdS/QCD~\cite{Forshaw:2012im}, LFH~\cite{Gurjar:2024wpq} are listed.
One can find that, our $\langle\xi^2\rangle_{2;\rho}^\parallel$ agrees well with the predictions by AdS/QCD~\cite{Forshaw:2012im} and QCD SRs~\cite{Bakulev:1998pf}, the $\langle\xi^n\rangle_{2;\phi}^\parallel$ at all orders by us are consistent with the values of the LQCD~\cite{Hua:2020gnw} within the error range, and for the $K^\ast$ meson, the deviation between our second-order $\xi$-moment result and the center value of DSE~\cite{Xu:2025hjf} and QCD SRs~\cite{Ball:2004rg} are less than $5\%$.

\begin{table}
\renewcommand{\arraystretch}{1.5}
\caption{Optimal fitting PLP model parameters for DA $\phi_{2;V}^\parallel(x,\mu) (V = \rho, K^\ast, \phi)$ and the values of the likelihood function and goodness of fit.}
\begin{tabular}{lcccccccccccccc}
\hline\hline
&&$\mu[\rm GeV]$     &&$\alpha_{2;V}^\parallel$     &&$\beta_{2;V}^\parallel$   &&$\chi^2_{\rm{min}}/n_d$  &&$P_{\chi^2_{\rm{min}}}$      \\[1pt]
\hline
$\rho$    &&1     &&$0.703$      &&$-$        &&0.121/4   &&$99.82\%$ \\[1pt]
$K^*$     &&1     &&$0.853$      &&$0.761$    &&1.942/8    &&$98.27\%$\\[1pt]
$\phi$    &&1     &&$0.895$      &&$-$        &&0.009/4      &&$99.99\%$\\[1pt]
\hline
$\rho$    &&1.4   &&$0.747$      &&$-$        &&0.131/4   &&$99.80\%$\\[1pt]
$K^*$     &&1.4   &&$0.872$      &&$0.787$    &&1.671/8    &&$98.94\%$ \\[1pt]
$\phi$    &&1.5   &&$0.912$       &&$-$       &&0.009/4   &&$99.99\%$  \\[1pt]
\hline
$\rho$    &&2     &&$0.775$      &&$-$         &&0.228/4  &&$99.40\%$\\[1pt]
$K^*$     &&2     &&$0.890$      &&$0.812$     &&1.632/8   &&$99.03\%$ \\[1pt]
$\phi$    &&2     &&$0.925$       &&$-$        &&0.016/4   &&$99.99\%$  \\[1pt]
\hline\hline
\end{tabular}
\label{tab:fit}
\end{table}

\begin{figure}[htp]
\centering
\includegraphics[width=0.45\textwidth]{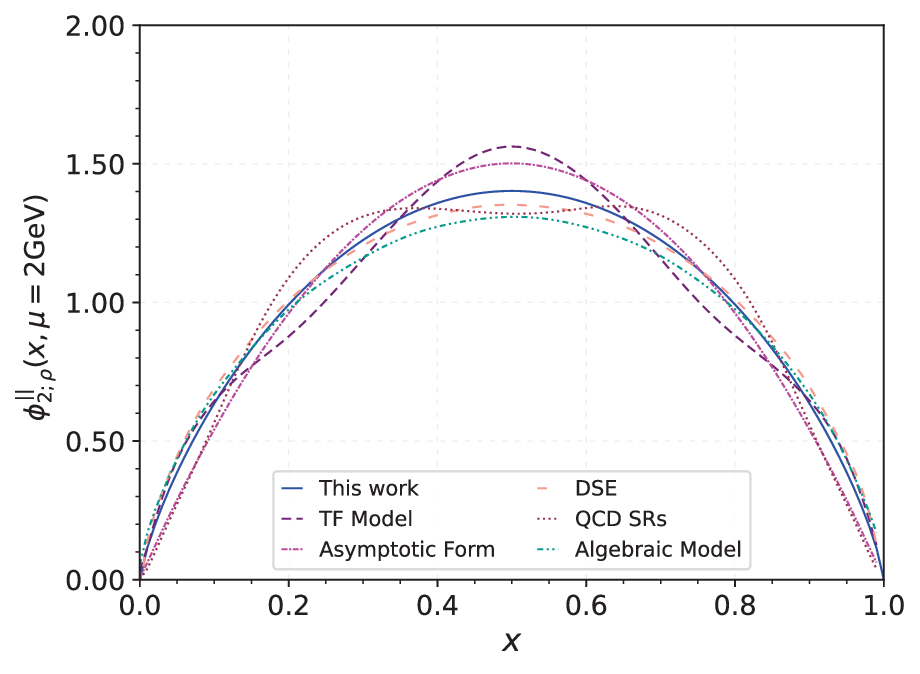}
\includegraphics[width=0.45\textwidth]{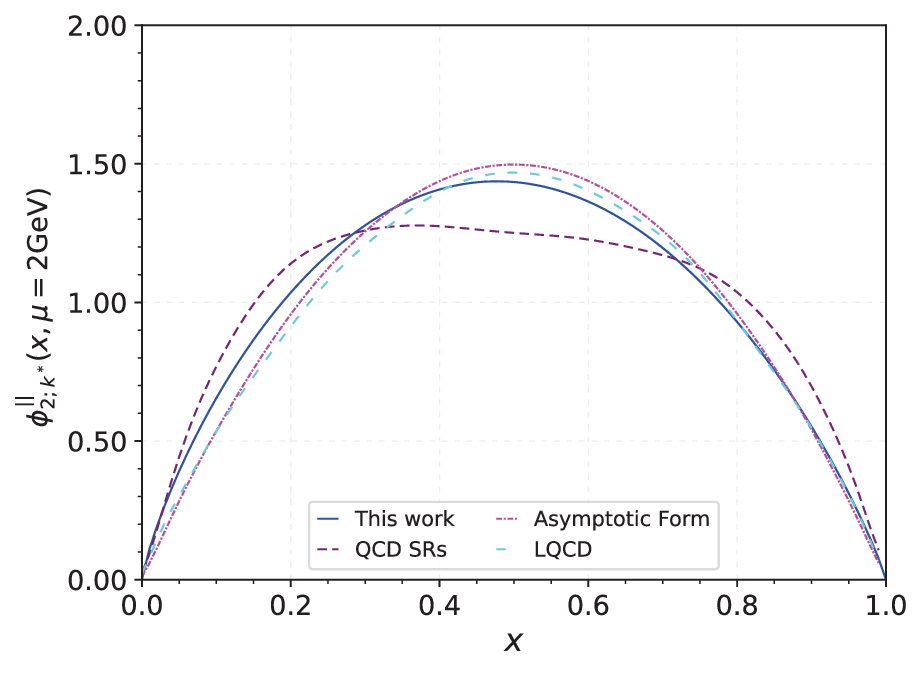}
\includegraphics[width=0.45\textwidth]{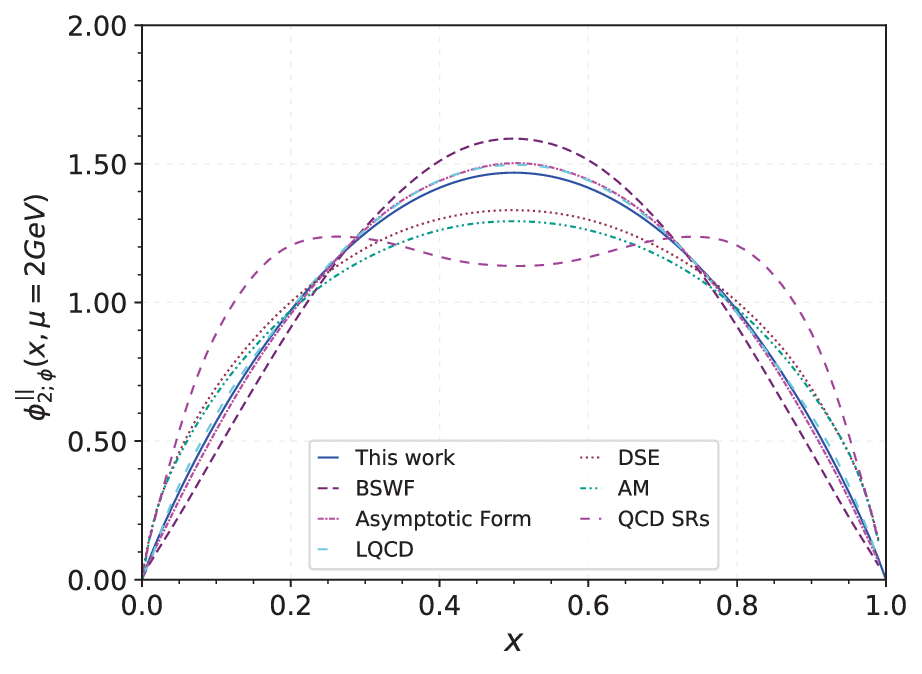}
\caption{Curves of the $\rho, K^\ast$ and $\phi$ meson leading-twist longitudinal DAs at $\mu=2~{\rm GeV}$. We also present the DAs by QCD SRs~\cite{Stefanis:2015qha, Ball:2007zt}, LQCD~\cite{Hua:2020gnw}, DSE~\cite{Gao:2014bca}, BSWF~\cite{Serna:2022yfp}, Algebraic Model~ \cite{Almeida-Zamora:2023rwg}, Asymptotic form~\cite{Lu:2007sg} and TF model~~\cite{Zhong:2023cyc} as a comparison.}
\label{fig:DAs}
\end{figure}

\subsection{Behaviors of vector meson leading-twist longitudinal DAs}
By fitting the values of $\langle\xi^n\rangle_{2;V}^\parallel$ exhibited in Tables~\ref{tab:xi_value1} and~\ref{tab:xi_value2} with the least squares method, the optimal fitting PLP model parameters for DAs $\phi_{2;V}(x,\mu) (V = \rho, K^\ast, \phi)$ at the typical scale $\mu = 1, 1.4, 1.5, 2~{\rm GeV}$ can be obtained. Those parameter values and the values of the likelihood function and goodness of fit are exhibited in Table~\ref{tab:fit}. Then, the behaviors of the $\rho, K^\ast, \phi$ leading-twist longitudinal DAs $\phi_{2;V}^\parallel(x,\mu)$ are determined, the corresponding curves of those DAs at scale $\mu = 2~{\rm GeV}$ are shown in Fig.~\ref{fig:DAs}. As a comparison, the curves of $\phi_{2;V}(x,\mu)$ predicted by various methods such as QCD SRs~\cite{Stefanis:2015qha, Ball:2007zt}, LQCD~\cite{Hua:2020gnw}, DSE~\cite{Gao:2014bca}, BSWF~\cite{Serna:2022yfp}, Algebraic model~ \cite{Almeida-Zamora:2023rwg}, Asymptotic form~\cite{Lu:2007sg} and the truncated form of Gegenbauer polynomial series (TF model)~\cite{Zhong:2023cyc} are also shown in Fig.~\ref{fig:DAs}. From Fig.~\ref{fig:DAs}, one can find that,
\begin{itemize}
\item For the $\rho$ meson, in the peak region, the results of this paper are higher than the predictions of the DSE~\cite{Gao:2014bca}, but lower than the Asymptotic form $\phi_{2;V}^\parallel(x,\mu = \infty) = 6x(1-x)$. In the endpoint region, our result is closer to the behavior of the TF model~~\cite{Zhong:2023cyc} and the Algebraic Model~\cite{Almeida-Zamora:2023rwg}.
\item For $K^\ast$ mesons, the result of this work is significantly different from that in Ref.~\cite{Ball:2007zt} within QCD SRs, but closer to LQCD result~\cite{Hua:2020gnw} and Asymptotic form in the overall shape. Affected by the $s$ quark mass effect, our DAs shows a slight asymmetry with a peak shift to $x<0.5$, which is consistent with the expectation of $SU_f(3)$ flavor symmetry breaking.
\item For $\phi$ mesons, there are obvious differences in the DA behaviors between our result and the prediction of early QCD SRs~\cite{Ball:2007zt}. In the peak region, the result of this paper are higher than the DSE~\cite{Gao:2014bca} and Algebraic model~\cite{Almeida-Zamora:2023rwg}, but lower than the prediction of the BSWF~\cite{Serna:2022yfp}. On the whole momentum fraction region, there is only a slight difference between the DA behavior given in this paper and the LQCD~\cite{Hua:2020gnw} and Asymptotic form.
\end{itemize}

In addition, the behaviors of $\phi_{2;V}^\parallel(x,\mu)$ with different scales exhibited in Table~\ref{tab:fit} show that our three DAs $\phi_{2;\rho}^\parallel(x,\mu)$, $\phi_{2;K^\ast}^\parallel(x,\mu)$ and $\phi_{2;\phi}^\parallel(x,\mu)$ tend to the Asymptotic form as the scale increases. This case is consistent with the expectation of the existing QCD conformal expansion.

\begin{table}
\renewcommand{\arraystretch}{1.2}
\caption{Fitting results of the PLP model for $\phi_{2;\rho}^\parallel$, $\phi_{2;\phi}^\parallel$ and $\phi_{2;K^\ast}^\parallel$ at $\mu=\sqrt{m_{D_{(s)}}^2-m_c^2}=1.4/1.5~\rm{GeV}$ obtained by fitting the first several non-zero $\xi$-moments listed in Tables~\ref{tab:xi_value1} and~\ref{tab:xi_value2}.}
\begin{tabular}{lccccccccccccccccccccccccc}
\hline\hline
   &&NS     &&2   &&3  &&4  &&5    \\[1pt]
\hline
$\phi_{2;\rho}^\parallel(x,\mu)$      &&$\alpha_{2;\rho}^\parallel$     &&0.755      &&0.756    &&0.755      &&0.747 \\[1pt]
              &&$\chi^2_{\rm{min}}/n_d$      &&0.001/1      &&0.001/2        &&0.001/3       &&0.131/4    \\[1pt]
              &&$P_{\chi^2_{\rm{min}}}$      &&$97.05\%$  &&$99.93\%$   &&$99.99\%$  &&$99.80\%$     \\[1pt]

\hline
$\phi_{2;\phi}^\parallel(x,\mu)$     &&$\alpha_{2;\phi}^\parallel$     &&0.907      &&0.912    &&0.915      &&0.912 \\[1pt]
              &&$\chi^2_{\rm{min}}/n_d$      &&0.001/1      &&0.004/2       &&0.006/3      &&0.009/4&&      \\[1pt]
              &&$P_{\chi^2_{\rm{min}}}$      &&$97.70\%$  &&$99.79\%$   &&$99.98\%$  &&$99.99\%$&&      \\[1pt]
 \hline
   &&NS     &&4   &&6  &&8  &&10    \\[1pt]
 \hline
$\phi_{2;K^\ast}^\parallel(x,\mu)$    &&$\alpha_{2;K^\ast}^\parallel$       &&$0.872$    &&0.873    &&0.873 &&0.872 \\[1pt]
              &&$\beta_{2;K^\ast}^\parallel$         &&$0.790$     &&0.789    &&0.788 &&0.787\\[1pt]
              &&$\chi^2_{\rm{min}}/n_d$           &&1.024/3       &&1.334/5      &&1.555/7 &&1.671/9\\[1pt]
             &&$P_{\chi^2_{\rm{min}}}$       &&$59.92\%$   &&$85.54\%$  &&$95.57\%$&&$98.94\%$\\[1pt]
\hline

\hline\hline
\end{tabular}
\label{xBta:1}
\end{table}

\begin{figure*}
    \begin{tabular}{cc}
        \includegraphics[width=0.3\textwidth]{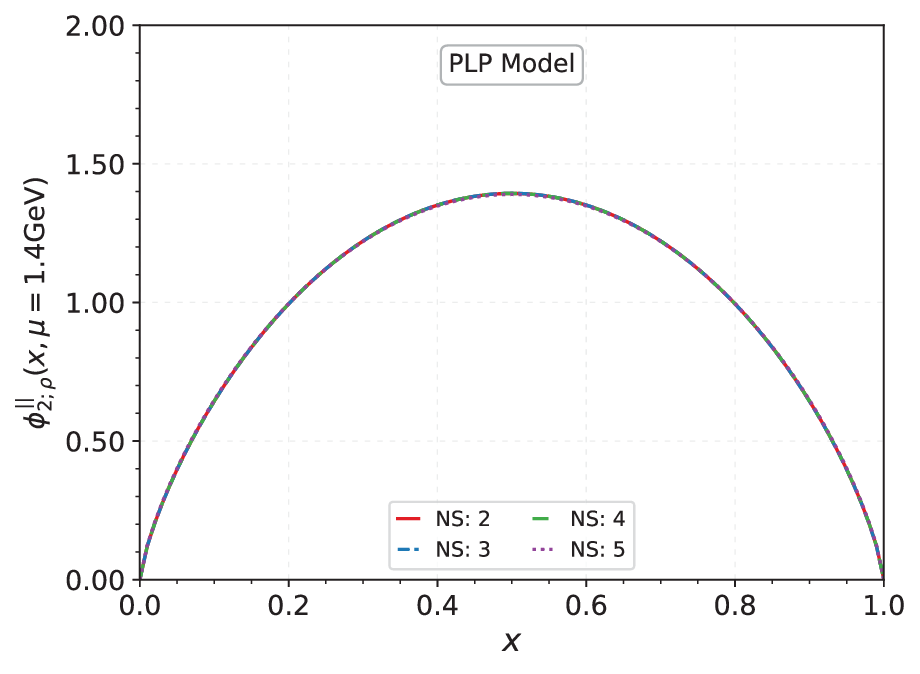} &
        \includegraphics[width=0.3\textwidth]{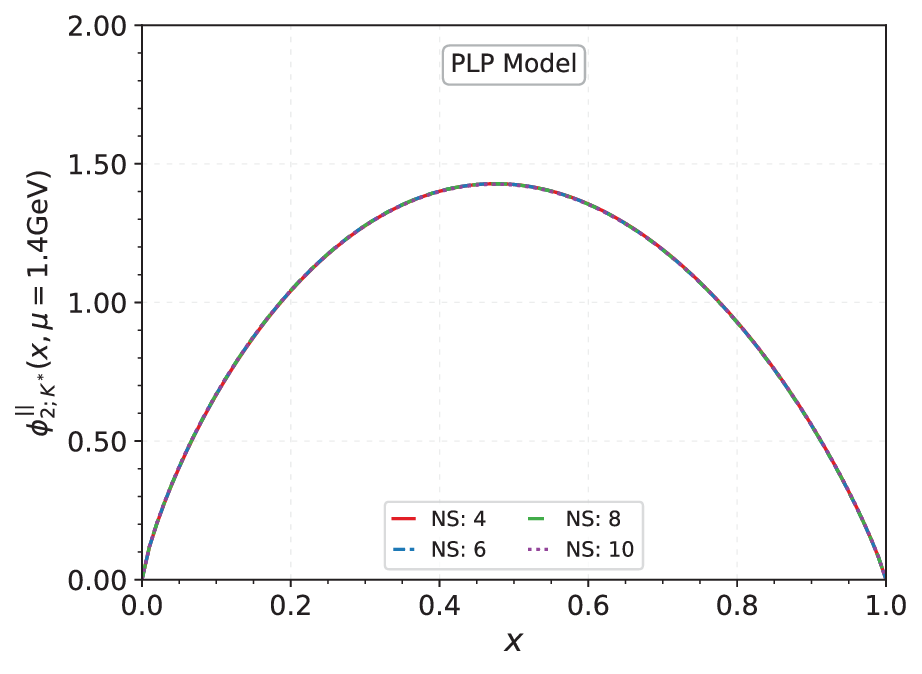}
        \includegraphics[width=0.3\textwidth]{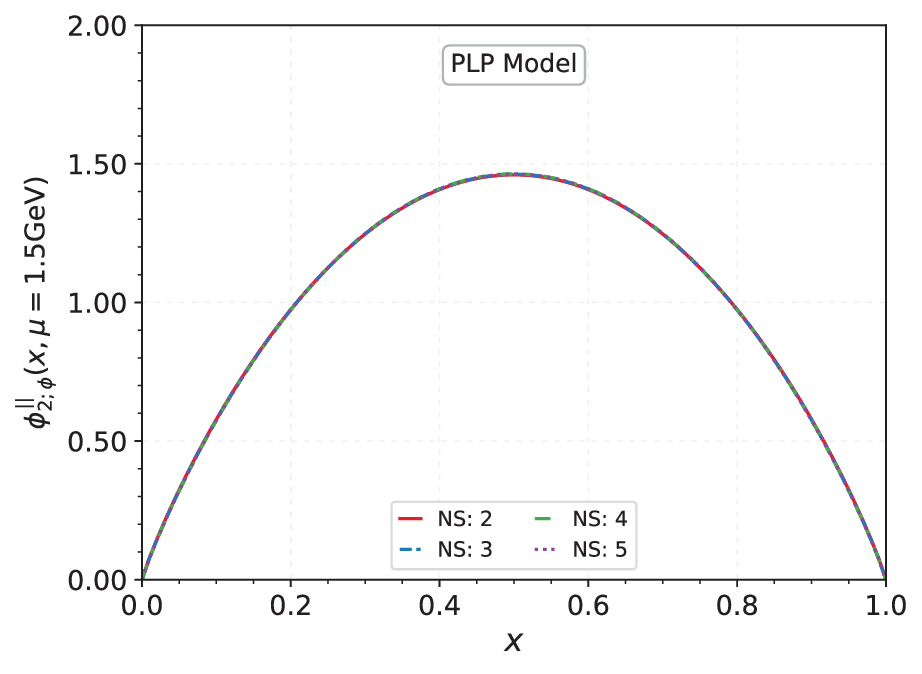}
        \\
       \includegraphics[width=0.3\textwidth]{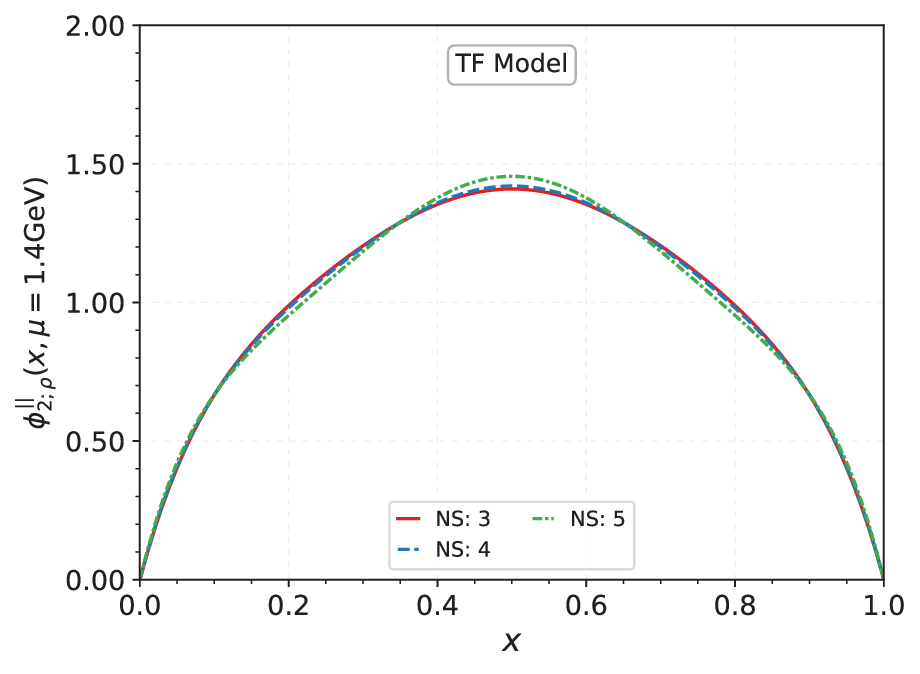}&
        \includegraphics[width=0.3\textwidth]{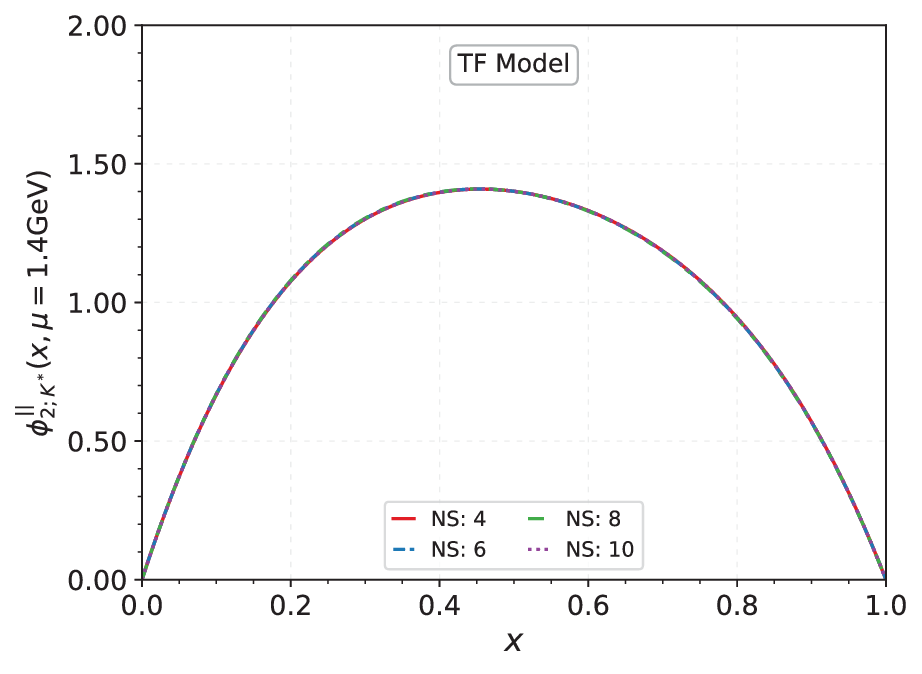}
        \includegraphics[width=0.3\textwidth]{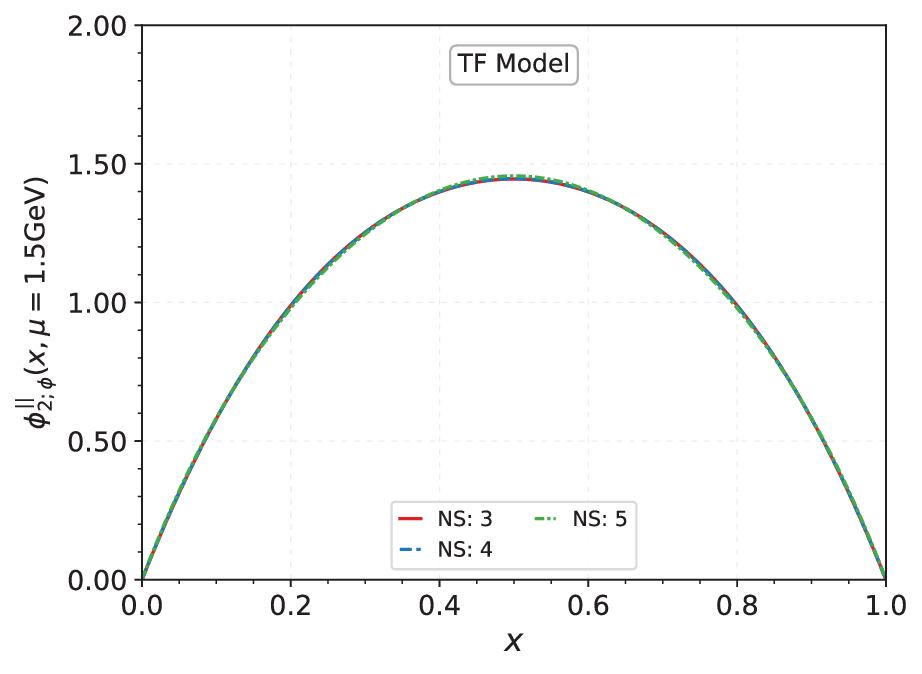}
        \\
         \includegraphics[width=0.3\textwidth]{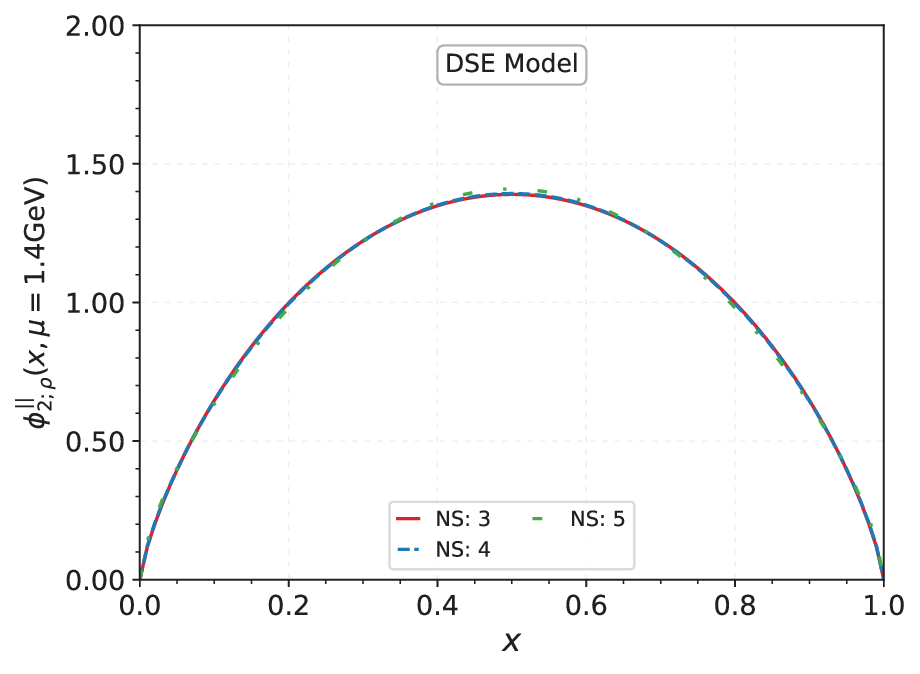}&
          \includegraphics[width=0.3\textwidth]{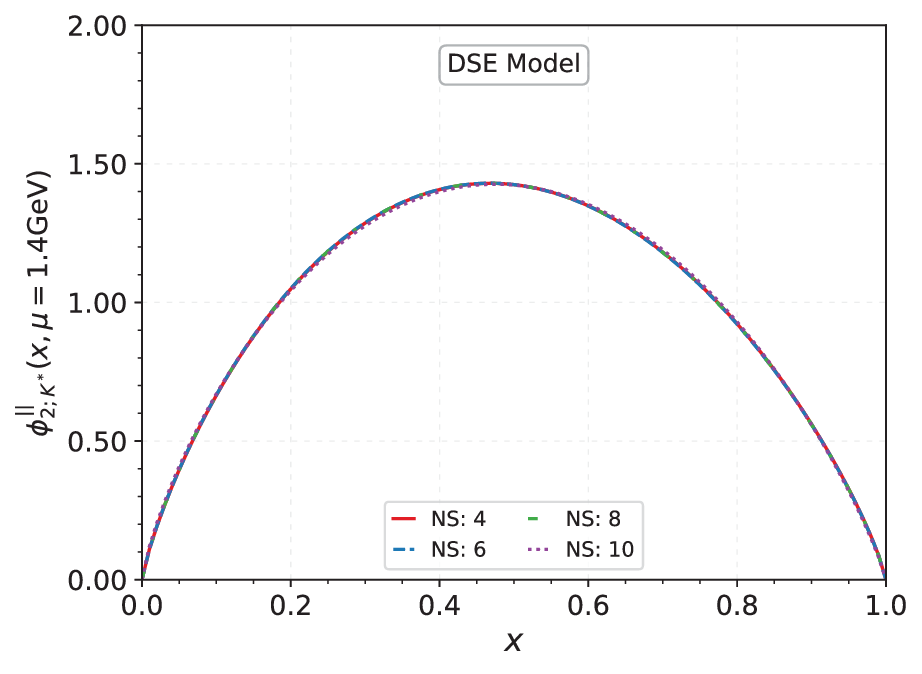}
        \includegraphics[width=0.3\textwidth]{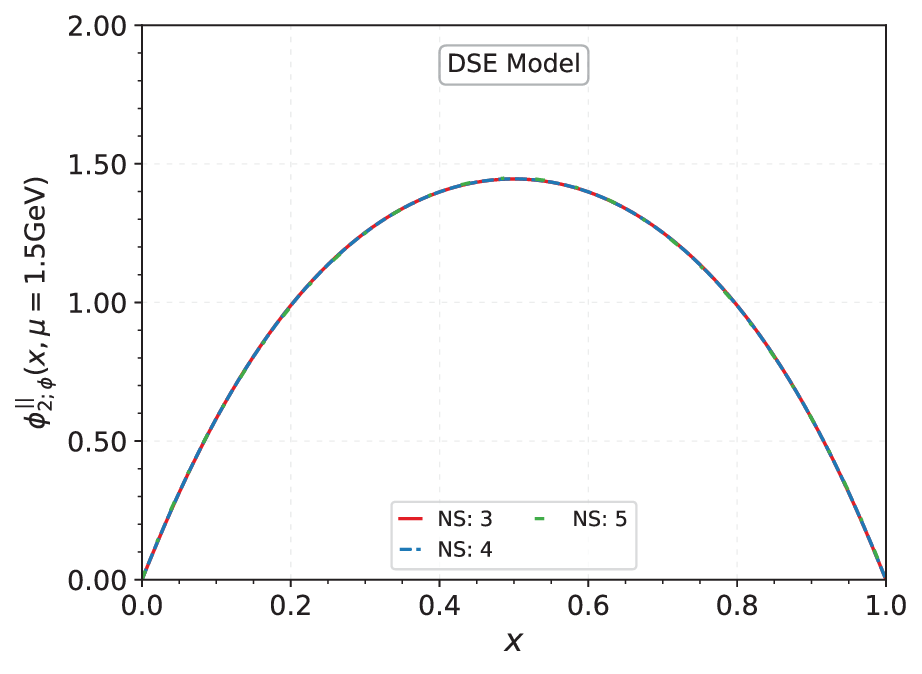}
    \end{tabular}
    \caption{Curves of $\phi_{2;\rho}^\parallel$, $\phi_{2;K^\ast}^\parallel$ and $\phi_{2;\phi}^\parallel$ at $\mu=\sqrt{m_{D_{(s)}}^2-m_c^2}=1.4/1.5~\rm{GeV}$ with different NS obtained by fitting with PLP, TF, and DSE models.}
    \label{xBfg:1}
\end{figure*}
On the other hand, it is meaningful to include different numbers of $\xi$-moments in the fitting to examine the stability and convergence of the obtained DAs. For this purpose, we set the values of the $\xi$-moments $\langle\xi^n\rangle_{2;\rho(\phi)}^\parallel$ at $\mu = 1.4/1.5~{\rm GeV}$ listed in Table~\ref{tab:xi_value1} to the form of the following four groups as the fitting samples,
\begin{align}
\rm{NS}:2, ~ &\big\{ \langle\xi^2\rangle_{2;\rho(\phi)}^\parallel, ~\langle\xi^4\rangle_{2;\rho(\phi)}^\parallel \big\} \nonumber\\
\rm{NS}:3, ~ &\big\{ \langle\xi^2\rangle_{2;\rho(\phi)}^\parallel, ~\langle\xi^4\rangle_{2;\rho(\phi)}^\parallel, ~\langle\xi^6\rangle_{2;\rho(\phi)}^\parallel \big\} \nonumber\\
\rm{NS}:4, ~ &\big\{ \langle\xi^2\rangle_{2;\rho(\phi)}^\parallel, ~\langle\xi^4\rangle_{2;\rho(\phi)}^\parallel, ~\langle\xi^6\rangle_{2;\rho(\phi)}^\parallel, ~\langle\xi^8\rangle_{2;\rho(\phi)}^\parallel \big\}\nonumber\\
\rm{NS}:5, ~ &\big\{ \langle\xi^2\rangle_{2;\rho(\phi)}^\parallel, ~\langle\xi^4\rangle_{2;\rho(\phi)}^\parallel, ~\langle\xi^6\rangle_{2;\rho(\phi)}^\parallel, ~\langle\xi^8\rangle_{2;\rho(\phi)}^\parallel, \nonumber\\
&\langle\xi^{10}\rangle_{2;\rho(\phi)}^\parallel \big\},
\end{align}
where NS is an abbreviation denoting the number of samples. Similarly, we take,
\begin{align}
\rm{NS}:4, ~ &\big\{ \langle\xi^1\rangle_{2;K^\ast}^\parallel, ~\langle\xi^2\rangle_{2;K^\ast}^\parallel, ~\langle\xi^3\rangle_{2;K^\ast}^\parallel, ~\langle\xi^4\rangle_{2;K^\ast}^\parallel \big\} \nonumber\\
\rm{NS}:6, ~ &\big\{ \langle\xi^1\rangle_{2;K^\ast}^\parallel, ~\langle\xi^2\rangle_{2;K^\ast}^\parallel, ~\langle\xi^3\rangle_{3;K^\ast}^\parallel, ~\langle\xi^4\rangle_{2;K^\ast}^\parallel, \nonumber\\ &\langle\xi^5\rangle_{2;K^\ast}^\parallel, ~\langle\xi^6\rangle_{2;K^\ast}^\parallel \big\} \nonumber\\
\rm{NS}:8, ~ &\big\{ \langle\xi^1\rangle_{2;K^\ast}^\parallel, ~\langle\xi^2\rangle_{2;K^\ast}^\parallel, ~\langle\xi^3\rangle_{2;K^\ast}^\parallel, ~\langle\xi^4\rangle_{2;K^\ast}^\parallel, \nonumber\\ &\langle\xi^5\rangle_{2;K^\ast}^\parallel, ~\langle\xi^6\rangle_{2;K^\ast}^\parallel, ~\langle\xi^7\rangle_{2;K^\ast}^\parallel, ~\langle\xi^8\rangle_{2;K^\ast}^\parallel \big\}\nonumber\\
\rm{NS}:10, ~ &\big\{ \langle\xi^1\rangle_{2;K^\ast}^\parallel, ~\langle\xi^2\rangle_{2;K^\ast}^\parallel, ~\langle\xi^3\rangle_{2;K^\ast}^\parallel, ~\langle\xi^4\rangle_{2;K^\ast}^\parallel, \nonumber\\ &\langle\xi^5\rangle_{2;K^\ast}^\parallel, ~\langle\xi^6\rangle_{2;K^\ast}^\parallel, ~\langle\xi^7\rangle_{2;K^\ast}^\parallel, ~\langle\xi^8\rangle_{2;K^\ast}^\parallel, \nonumber\\ &\langle\xi^9\rangle_{2;K^\ast}^\parallel, ~\langle\xi^{10}\rangle_{2;K^\ast}^\parallel \big\},
\end{align}
for the values of $\langle\xi^n\rangle_{2;K^\ast}^\parallel$ at $\mu = 1.4~{\rm GeV}$ listed in Table~\ref{tab:xi_value2}. The fitting results for PLP model are exhibited in Table~\ref{xBta:1}, and the corresponding DA curves are shown in Fig.~\ref{xBfg:1}. One can find that,
\begin{itemize}
\item For $\phi_{2;\rho}^\parallel$ and $\phi_{2;\phi}^\parallel$, the difference in fitting parameter values between different $\rm NS$ is very small. For the case of ${\rm NS} \geq 4$, the goodness of fit remains above $99\%$.
\item Similar to the situation with $\phi_{2;\rho(\phi)}^\parallel$, the difference in fitting parameter values between different $\rm NS$ is also very small for $\phi_{2;K^\ast}^\parallel$. However, the goodness of fit increases with the increase of ${\rm NS}$.
\item The curves of $\phi_{2;\rho}^\parallel$, $\phi_{2;K^\ast}^\parallel$ and $\phi_{2;\phi}^\parallel$ for different ${\rm NS}$ almost completely overlap, respectively.
\end{itemize}
The above results are different from the case of the pion leading-twist DA $\phi_{2;\pi}$. In our previous work~\cite{Zhong:2022lmn}, a similar analysis for $\phi_{2;\pi}$ was performed, and we found that the obtained DA behavior gradually becomes fully stable with the inclusion of an increasing number of $\xi$-moments. On the contrary, for $\phi_{2;\rho}^\parallel$, $\phi_{2;K^\ast}^\parallel$ and $\phi_{2;\phi}^\parallel$, the fitted parameters and the DA shapes show only small change with the variation of NS.

In order to further test whether this result depends on specific model assumptions, we introduce two commonly used phenomenological models for cross comparison, namely TF model $\phi^\parallel_{2;V}(x,\mu) = 6x(1-x) [1 + a_{2;V}^1 C_1^{3/2}(2x-1) + a_{2;V}^2 C_2^{3/2}(2x-1)]$~\cite{Ball:1996tb} and DSE model $\phi^\parallel_{2;V}(x,\mu) = \mathcal{N} [x(1-x)]^{\alpha\_} [1 + a_{2;V}^1 C_1^\alpha(2x-1) + a_{2;V}^2 C_2^\alpha(2x-1)]$~\cite{Gao:2014bca}, where $a_{2;V}^n$ is the Gegenbauer moments and is used as a fitting parameters here ($a_{2;\rho(\phi)}^1 = 0$ due to the isospin symmetry), $C_n^{3/2}(2x-1)$ and $C_n^\alpha(2x-1)$ are the $n$th order Gegenbauer polynomials, $\alpha\_ = \alpha - 1/2$, the normalization constant $\mathcal{N} = \Gamma(2\alpha + 1) / [\Gamma (\alpha + 1/2)]^2$ with the normalization condition of the vector meson leading-twist longitudinal DA. The corresponding results are also shown in Fig.~\ref{xBfg:1}. One can find that, similar to the PLP model, the DA shapes fitted using the TF model and DSE model still have a very weak dependence on NS.

The weak NS dependence may be related to the fact that these DA are already close to the Asymptotic form~\cite{Lu:2007sg}. The numerical results also show that higher-order moments do not significantly change the reconstructed DA shapes. However, their calculation remains useful. These moments provide additional nonperturbative constraints and allow a quantitative test of the internal consistency and stability of the obtained $\phi_{2;V}^\parallel$. In the subsequent calculations for $D_{(s)}\to V$ semi-leptonic decays, we still use the $\phi_{2;V}^\parallel$ determined by the PLP model combined with the fitting parameters in Table~\ref{tab:fit} as the input parameters.

\subsection{$D_{(s)}\to V$ semi-leptonic decays}
To calculate the $D_{(s)}\to V$ TFFs, the renormalization scale is taken as $\mu_{\rm IR} = \sqrt{m_{D_{(s)}}^2 - m_c^2} \approx 1.4/1.5~\rm GeV$. The bound state parameters are $m_D = 1.869 \pm 0.05 \rm~GeV$, $m_{D_s} = 1.968 \pm 0.07 \rm~ GeV$~\cite{ParticleDataGroup:2024cfk}, $f_D = 0.212 \pm 0.007 \rm~GeV$ \cite{Bazavov:2017lyh}, $f_{D_s} = 0.257 \pm 0.004 \rm~GeV$ \cite{Duplancic:2015zna}. In addition, the continuum threshold and Borel parameter are two essential inputs. The former is used to separate the ground state and excited state contributions, and the latter is used to suppress the influence of high excited states and continuous states and then ensure the convergence of OPE~\cite{Ball:2005vx}. Following the those criteria, we take the continuum threshold as the value closest to the squared mass of the first excited state of the initial state charm meson, i.e., $s_D = 6.76 \pm 0.50 \rm~GeV^2$, $s_{D_s} = 7.2 \pm 0.50 \rm~GeV^2$. We take the Borel parameters as $M_{V;\rho}^2 = 1.89 \pm 0.06 \rm~GeV^2$, $M_{A_1;\rho}^2 = 1.20 \pm 0.05 \rm~GeV^2$, $M_{A_2;\rho}^2 = 1.13 \pm 0.03 \rm~GeV^2$, $M_{V;K^\ast}^2 = 2.19 \pm 0.04 \rm~GeV^2$, $M_{A_1;K^\ast}^2 = 1.23 \pm 0.05 \rm~GeV^2$, $M_{A_2;K^\ast}^2 = 1.30 \pm 0.02 \rm~GeV^2$, and $M_{V;\phi}^2 = 2.65 \pm 0.11 \rm~GeV^2$, $M_{A_1;\phi}^2 = 2.05 \pm 0.20 \rm~GeV^2$, $M_{A_2;\phi}^2 = 1.84 \pm 0.05 \rm~GeV^2$, respectively.

\begin{table}[h]
\renewcommand{\arraystretch}{1.2}
\scalebox{0.6}
\footnotesize
\caption{Predictions for the $D\to\rho$ TFFs $V^{D\to\rho}(0)$, $A^{D\to\rho}_{1,2}(0)$, at the large recoil point $q^2=0$. For comparison, other theoretical predictions are also provided.}
\label{table:ttf1}
\begin{tabular}{lllll}
\hline\hline
~~~~~~~~~~~~~~~~~~& $A_1^{D\to\rho}(0)$               &$A_2^{D\to\rho}(0)$                       &$V^{D\to\rho}(0)$&  \\ \hline
This work   & $0.528^{+0.039}_{-0.036}$    & $0.478^{+0.045}_{-0.042}$      &$0.806^{+0.053}_{-0.052}$& \\
CLEO'13 \cite{CLEO:2011ab}   & $0.56(1)^{+0.02}_{-0.03}$    & $0.47(6)(4)$             &$0.84(9)^{+0.05}_{-0.06}$ &\\
LQCD~\cite{Lubicz:1991bi}       & $0.45(4)$                     & $0.02(26)$                      &$0.78(12)$& \\
LQCD~\cite{Bernard:1991bz}       & $0.65(15)^{+0.24}_{-0.23}$    & $0.59(31)^{+0.28}_{-0.25}$     &$1.07(49)(35)$& \\
LQCD~\cite{Flynn:1997ca}          &$0.65(7)$                      &$0.55(10)$                      &$1.1(2)$\\
LEChQM~\cite{Palmer:2013yia}      & $0.56$                        & $0.47$                           &$0.84$& \\
3PSRs\cite{Ball:1993tp}         & $0.5(2)$                     & $0.4(1)$                &$1.0(2)$& \\
HQETF~\cite{Wang:2002zba}       & $0.57(8)$                     & $0.52(7)$               &$0.72(10)$& \\
RHOPM~\cite{Wirbel:1985ji}      & $0.78$                        & $0.92$                            &$1.23$& \\
LCSRs~\cite{Wu:2006rd}      & $0.599^{+0.035}_{-0.030}$     & $0.372^{+0.026}_{-0.031}$        &$0.801^{+0.044}_{-0.036}$  & \\
LFQM~\cite{Verma:2011yw}       & $0.60(1)$                      & $0.47(0)$                      &$0.88(3)$& \\
HM$\chi$T~\cite{Fajfer:2005ug}        & $0.61$                        & $0.31$                          &$1.05$& \\
QM-I~\cite{Isgur:1988gb}        & $0.59$                        & $0.23$                          &$1.34$& \\
QM-II~\cite{Melikhov:2000yu}       & $0.59$                        & $0.49$                          &$0.90$& \\
\hline\hline
\end{tabular}
\end{table}
\begin{table}[h]
\renewcommand{\arraystretch}{1.2}
\scalebox{0.6}
\footnotesize
\caption{Predictions for the $D\to K^\ast$ TFFs $V^{D\to K^\ast}(0)$, $A^{D\to K^\ast}_{1,2}(0)$, at the large recoil point $q^2=0$. For comparison, other theoretical predictions are also provided.}
\label{table:ttf2}
\begin{tabular}{lllll}
\hline\hline
~~~~~~~~~~~~~~~~~~& $A^{D\to K^\ast}_1(0)$                  &$A^{D\to K^\ast}_2(0)$                       &$V^{D\to K^\ast}(0)$&  \\ \hline
This work         & $0.566^{+0.034}_{-0.032}$  & $0.510^{+0.030}_{-0.028}$     &$0.876^{+0.042}_{-0.039}$& \\[2pt]
LQCD~\cite{APE:1994kxx}    & $0.67\pm0.03$      & $0.49\pm0.34$                &$1.08\pm0.22$& \\[2pt]
LQCD~\cite{Lubicz:1991bi}   & $0.53\pm0.11$      & $0.19\pm0.21$               &$0.86\pm0.03$& \\[2pt]
LQCD~\cite{Lubicz:1990pi}    & $0.52\pm0.07$        & $0.35\pm0.05$           &$0.85\pm0.08$& \\[2pt]
LCSRs~\cite{Wu:2006rd}        & $0.571^{+0.024}_{-0.026}$ & $0.345^{+0.034}_{-0.037}$  &$0.791^{+0.032}_{-0.027}$& \\[2pt]
QCD SRs~\cite{Ball:1991bs}        & $0.5\pm0.15$       & $0.6\pm0.15$            &$1.1\pm0.25$& \\[2pt]
LFQM~\cite{Verma:2011yw}       & $0.72$             & $0.60$                 &$0.98$        & \\[2pt]
LFQM~\cite{Demchuk:1997uz}     & $0.633$            & $0.464$                &$0.777$& \\[2pt]
QM-II~\cite{Melikhov:2000yu}    & $0.66$            & $0.49$                  &$1.03$& \\[2pt]
HM$\chi$T~\cite{Fajfer:2005ug}        & $0.62$           & $0.31$                 &$0.99$& \\[2pt]
RQM~\cite{Faustov:2019mqr}        & $0.608$           & $0.520$              &$0.927$& \\[2pt]
\hline\hline
\end{tabular}
\end{table}
\begin{table}[htp]
\renewcommand{\arraystretch}{1.2}
\scalebox{0.6}
\footnotesize
\caption{Predictions for the $D_s\to\phi$ TFFs $V^{D_s\to\phi}(0)$, $A^{D_s\to\phi}_{1,2}(0)$, at the large recoil point $q^2=0$. For comparison, other theoretical predictions are also provided.}
\label{table:ttf3}
\begin{tabular}{lllll}
\hline\hline
~~~~~~~~~~~~~~~~~~& $A^{D_s\to\phi}_1(0)$               &$A^{D_s\to\phi}_2(0)$                             &$V^{D_s\to\phi}(0)$&  \\ \hline
This work        & $0.532^{+0.037}_{-0.028} $      & $0.439^{+0.067}_{-0.060}$   &$0.887^{+0.036}_{-0.035}$ \\[2pt]
LQCD~\cite{Gill:2001jp}    & $0.63(2) $       & $0.62(5)$     &$0.85(4)$ \\[2pt]
LQCD~\cite{Donald:2011ff}  & $0.594(22) $     & $0.401(80)$      &$0.903(67)$ \\[2pt]
LQCD~\cite{Donald:2013pea} & $0.615(24)$     & $0.457(78)$      &$1.059(124)$ \\[2pt]
CCQM~\cite{Ivanov:2019nqd}         & $0.68 $            & $0.67$         &$0.91$ \\[2pt]
CQM~\cite{Melikhov:2000yu}         & $0.64$& $0.47$&$1.10$ \\[2pt]
HM$\chi$T~\cite{Fajfer:2005ug} & $0.61 $& $0.32$&$1.10$ \\[2pt]
3PSR~\cite{Du:2003ja} & $0.55(15)$& $0.0.59(11)$&$1.21(33)$ \\[2pt]
LFQM~\cite{Wang:2008ci}& $0.61$& $0.58$&$0.91$ \\[2pt]
LFQM~\cite{Verma:2011yw}& $0.69 $& $0.57$&$0.98$ \\[2pt]
LCSRs~\cite{Aliev:2004vf} & $0.54(9) $& $0.57(9)$&$0.70(10)$ \\[2pt]
RQM~\cite{Faustov:2019mqr} & $0.643 $& $0.492$&$0.999$ \\[2pt]
\hline\hline
\end{tabular}
\end{table}

\begin{table}[h]
\renewcommand{\arraystretch}{1.2}
\scalebox{0.6}
\footnotesize
\caption{$D_{(s)}\to V$ TFFs at $q^2=0$ with $\phi_{2;\rho(\phi)}^\parallel$ corresponding to $\rm{NS} = 2, 5$ and $\phi_{2;K^\ast}^\parallel$ corresponding to $\rm{NS} = 4, 10$ from Table~\ref{xBta:1}, respectively.}
\label{tab:NS}
\begin{tabular}{lllll}
\hline\hline
~~~~~~~~~~~~~~~~~~& $A_1^{D\to\rho}(0)$               &$A_2^{D\to\rho}(0)$                       &$V^{D\to\rho}(0)$&  \\ \hline
   NS: 5& $0.528^{+0.039}_{-0.036}$    & $0.478^{+0.045}_{-0.042}$      &$0.806^{+0.053}_{-0.052}$& \\
   NS: 2& $0.528^{+0.039}_{-0.037}$    & $0.479^{+0.045}_{-0.042}$      &$0.806^{+0.053}_{-0.052}$& \\
\hline
~~~~~~~~~~~~~~~~~~& $A_1^{D\to K^*}(0)$     &$A_2^{D\to K^*}(0)$                       &$V^{D\to K^*}(0)$&  \\ \hline
   NS: 10& $0.566^{+0.034}_{-0.032}$    & $0.510^{+0.030}_{-0.028}$      &$0.876^{+0.042}_{-0.039}$& \\
   NS: 4& $0.566^{+0.034}_{-0.032}$    & $0.509^{+0.030}_{-0.029}$      &$0.876^{+0.041}_{-0.040}$& \\
\hline
~~~~~~~~~~~~~~~~~~& $A_1^{D_s\to\phi}(0)$               &$A_2^{D_s\to\phi}(0)$                       &$V^{D_s\to\phi}(0)$&  \\ \hline
   NS: 5& $0.532^{+0.037}_{-0.028}$    & $0.439^{+0.067}_{-0.060}$      &$0.887^{+0.036}_{-0.035}$& \\
   NS: 2& $0.532^{+0.037}_{-0.028}$     & $0.439^{+0.067}_{-0.060}$       &$0.887^{+0.036}_{-0.035}$& \\
\hline\hline
\end{tabular}
\end{table}

Then, we calculate the TFFs $A_{1,2}(q^2)$, $V(q^2)$ for $D\to(\rho, K^\ast)$ and $D_s\to\phi$ semi-leptonic decays. The corresponding values at the large recoil point ($q^2=0\rm~GeV$) are exhibited in Tables~\ref{table:ttf1}, \ref{table:ttf2}, and~\ref{table:ttf3}, respectively\footnote{In Tables~\ref{table:ttf1}, \ref{table:ttf2}, and~\ref{table:ttf3}, and the subsequent calculations and discussions, we added the superscript $D\to(\rho, K^\ast)$ and $D_s\to\phi$ to distinguish the TFFs $A_{1,2}(q^2)$, $V(q^2)$ of different semi-leptonic decay processes.}. As a comparison, we also list the results from CLEO collaboration~\cite{CLEO:2011ab} and various theoretical methods such as LQCD~\cite{Lubicz:1991bi, Bernard:1991bz, Flynn:1997ca, Lubicz:1990pi, Gill:2001jp, Donald:2011ff, Donald:2013pea},
LEChQM~\cite{Palmer:2013yia}, 3PSRs~\cite{Du:2003ja,Ball:1993tp}, HQETF~\cite{Wang:2002zba}, LCSRs~\cite{Wu:2006rd,Aliev:2004vf}, QM~\cite{Isgur:1988gb, Melikhov:2000yu}, LFQM~\cite{Verma:2011yw,
Demchuk:1997uz, Wang:2008ci}, HM$\chi$T~\cite{Fajfer:2005ug}, RQM~\cite{Faustov:2019mqr}, QCD SRs~\cite{Ball:1991bs}, CCQM~\cite{Ivanov:2019nqd}, CQM~\cite{Melikhov:2000yu}, RHOPM~\cite{Wirbel:1985ji} in those tables. As shown in the tables~\ref{table:ttf1},~\ref{table:ttf2} and~\ref{table:ttf3}, the relative uncertainties of the TFFs $A_{1,2}^{D_{(s)}\to V}(0)$ and $V^{D_{(s)}\to V}(0)$ at the large recoil point are between $5\%$ and $15\%$ and which come from the squared average for the input parameters such as decay constant, meson mass, continuum threshold and Borel parameter, etc. Within errors, our results for $A_{1,2}^{D\to\rho}(0)$ and $V^{D\to\rho}(0)$ are consistent with the CLEO 2013 data~\cite{CLEO:2011ab} and the LEChQM predictions~\cite{Palmer:2013yia}, $A_{1,2}^{D\to\rho}(0)$ also agree with the 3PSRs results~\cite{Ball:1993tp}. The $A_1^{D\to K^\ast}(0)$ and $V^{D\to K^\ast}(0)$ by us are consistent with LQCD values~\cite{Lubicz:1991bi,Lubicz:1990pi} within errors. Our prediction for $V^{D_s\to\phi}(0)$ is also consistent with LQCD predictions~\cite{Gill:2001jp,Donald:2011ff}.

It is meaningful to further investigate the impact of fitting $\phi_{2;V}^\parallel$ with different NS on predictions of TFFs. By keeping all other input parameters fixed, we compare the TFFs obtained with $\rm{NS} = 2$ and $\rm{NS} = 5$ for $D_{(s)}\to(\rho, \phi)$, and compare the TFFs obtained with $\rm{NS} = 4$ and $\rm{NS} = 10$ for $D\to K^\ast$. The results are listed in Table~\ref{tab:NS}. For all three transitions, the differences between the two sets of TFFs are almost negligible. This is consistent with the fact that the corresponding fitting DA curves almost coincide. This comparison further demonstrates the stability of our DAs from the perspective of form factor predictions.

We also give the TFFs ratios at the large recoil point $(q^2=0)$ with $r_V=V(0)/A_1(0)$, $r_2=A_2(0)/A_1(0)$,
\begin{align}
&r_V^{D\to\rho} = 1.527^{+0.012}_{-0.006}, & &r_2^{D\to\rho} = 0.905^{+0.017}_{-0.019}, \nonumber \\
&r_V^{D\to K^*} = 1.548^{+0.017}_{-0.017}, & &r_2^{D\to K^*} = 0.901^{+0.012}_{-0.012}, \nonumber \\
&r_V^{D\to\phi} = 1.667^{+0.045}_{-0.023}, & &r_2^{D\to\phi} = 0.825^{+0.064}_{-0.073}. \nonumber
\end{align}
Within errors, our $r_V^{D\to\rho}$ predictions agree with the BESIII~\cite{BESIII:2024lxg} measurements value of $r_V=1.548\pm0.079\pm0.041$, and the $r_V^{D\to\phi}$ predictions also agree with the BESIII~\cite{BESIII:2023opt} measurements $r_V=1.58\pm0.17\pm0.02$.

\begin{figure*}[htp]
\centering
\includegraphics[width=0.3\textwidth]{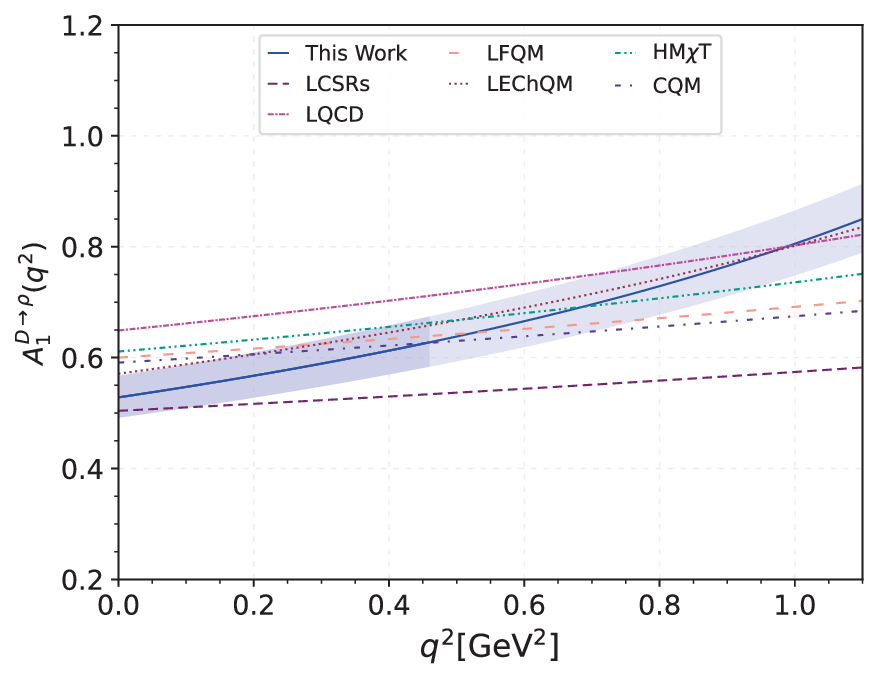} \hspace{1em}
\includegraphics[width=0.3\textwidth]{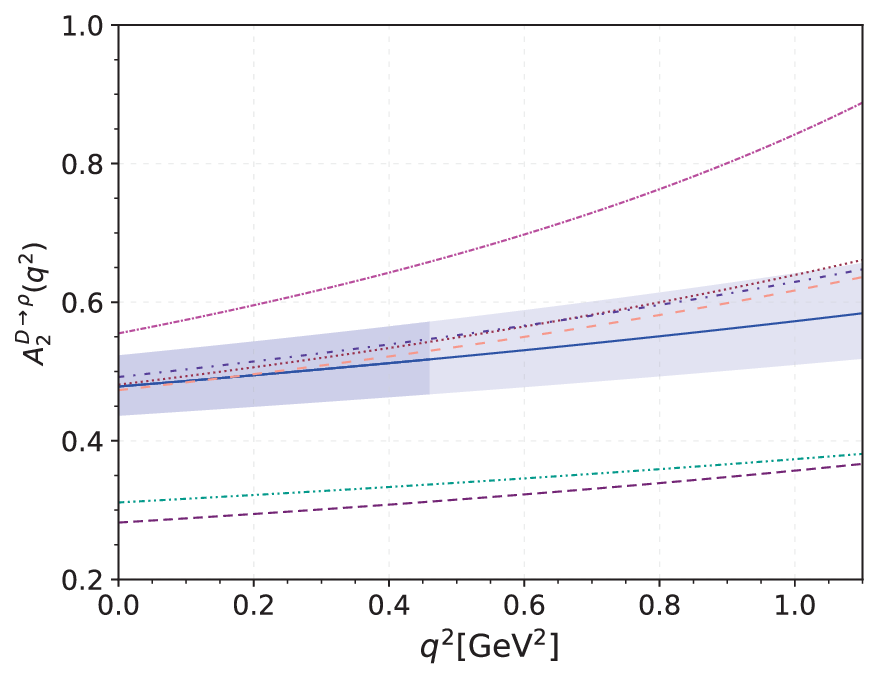} \hspace{1em}
\includegraphics[width=0.3\textwidth]{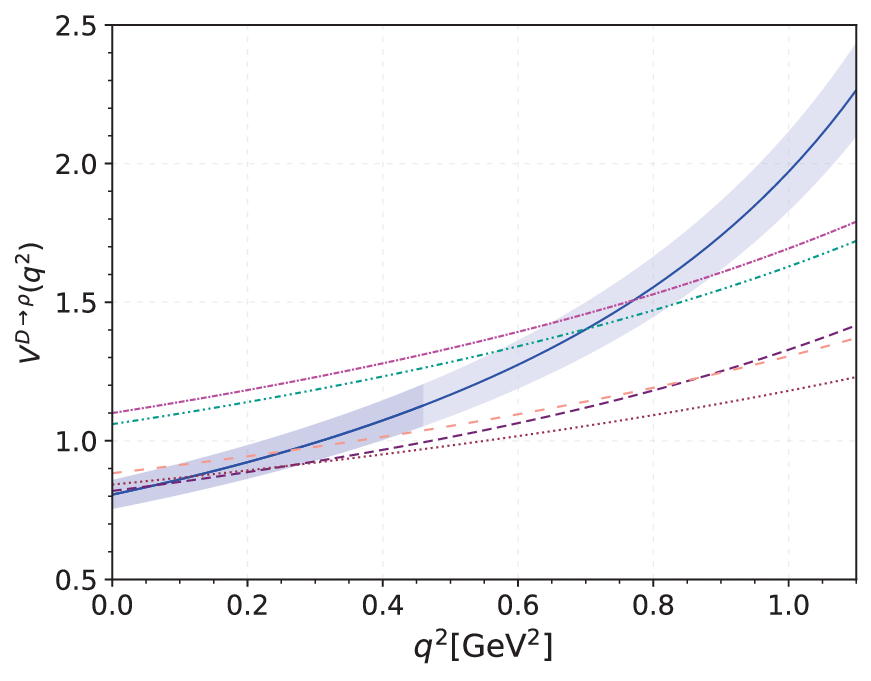} \\[5pt]
\includegraphics[width=0.3\textwidth]{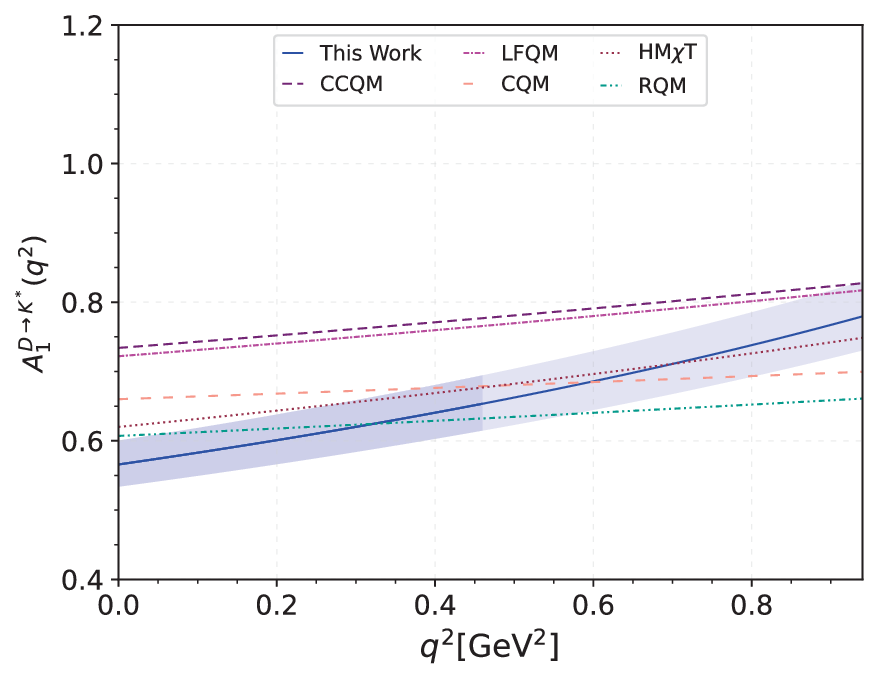} \hspace{1em}
\includegraphics[width=0.3\textwidth]{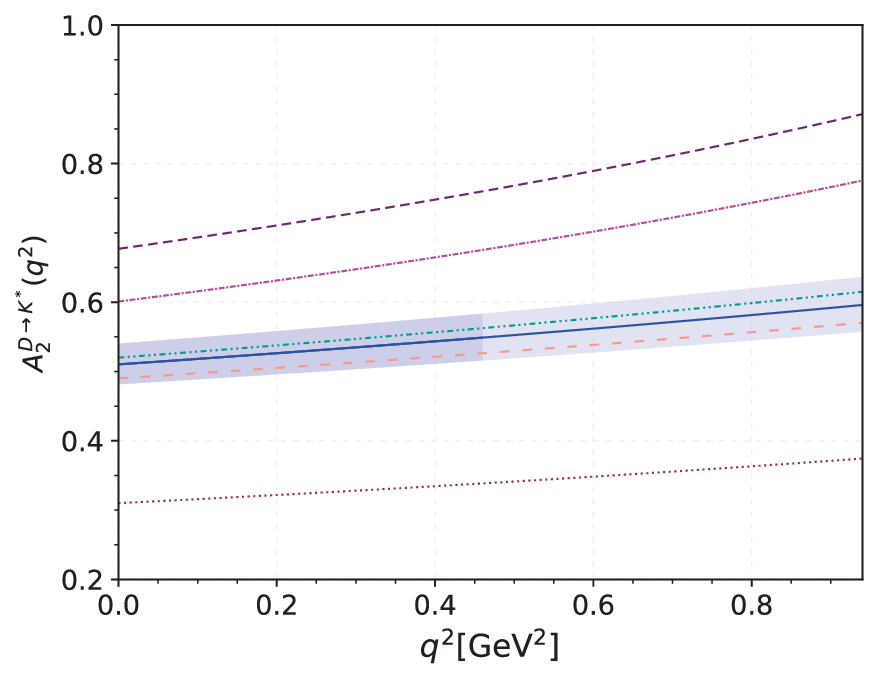} \hspace{1em}
\includegraphics[width=0.3\textwidth]{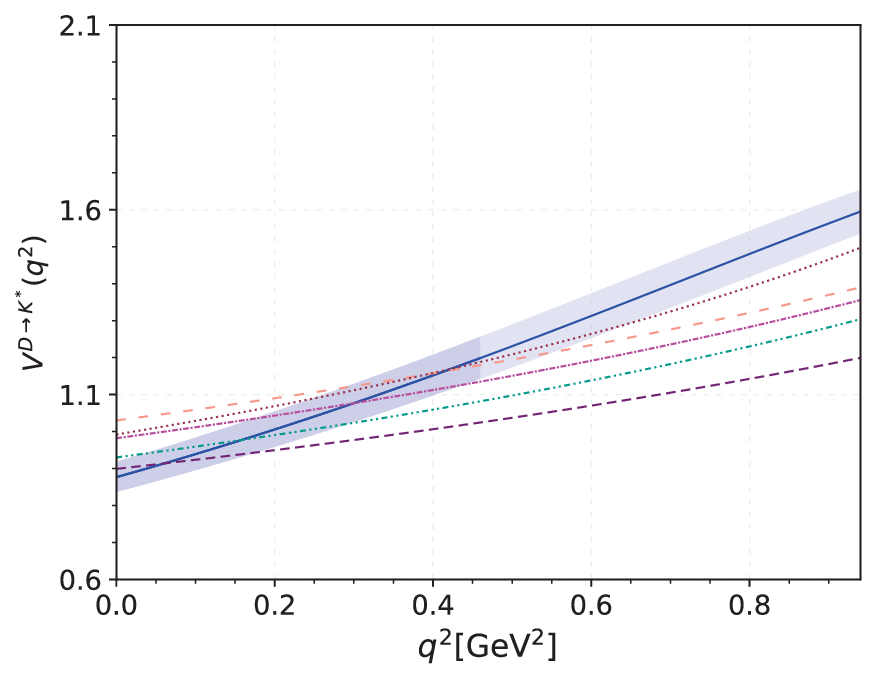} \\[5pt]
\includegraphics[width=0.3\textwidth]{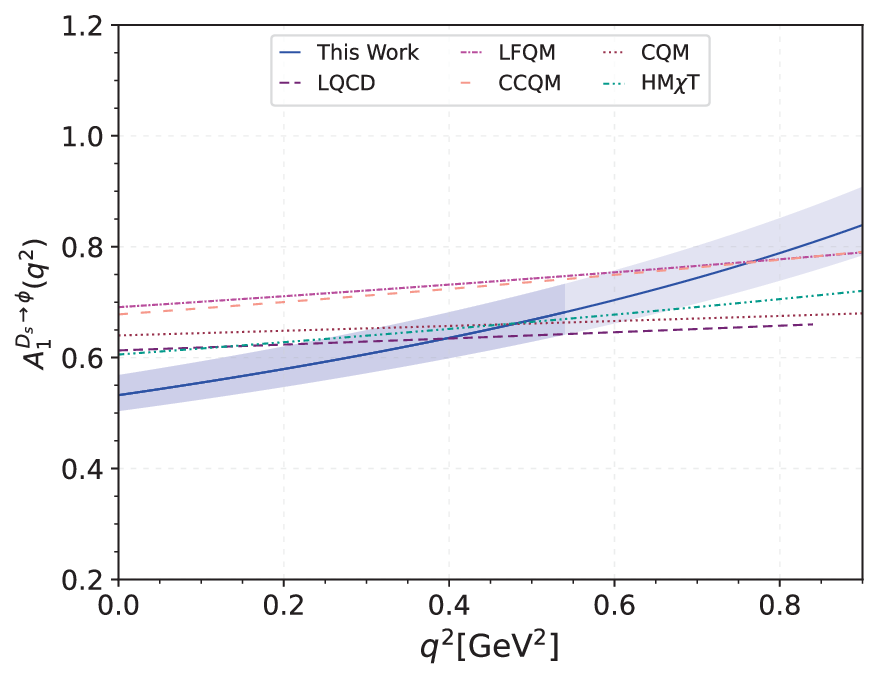} \hspace{1em}
\includegraphics[width=0.3\textwidth]{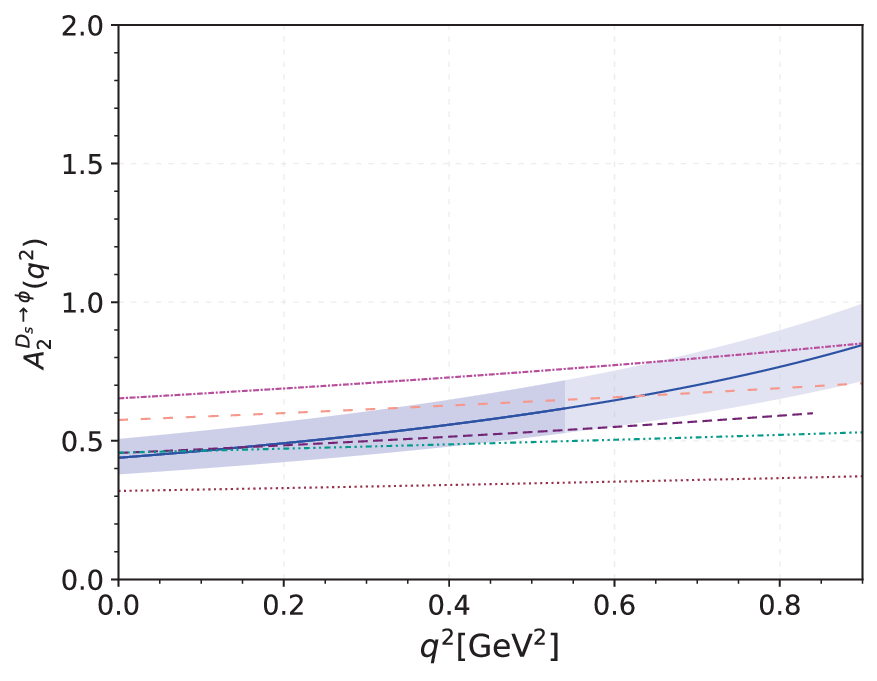} \hspace{1em}
\includegraphics[width=0.3\textwidth]{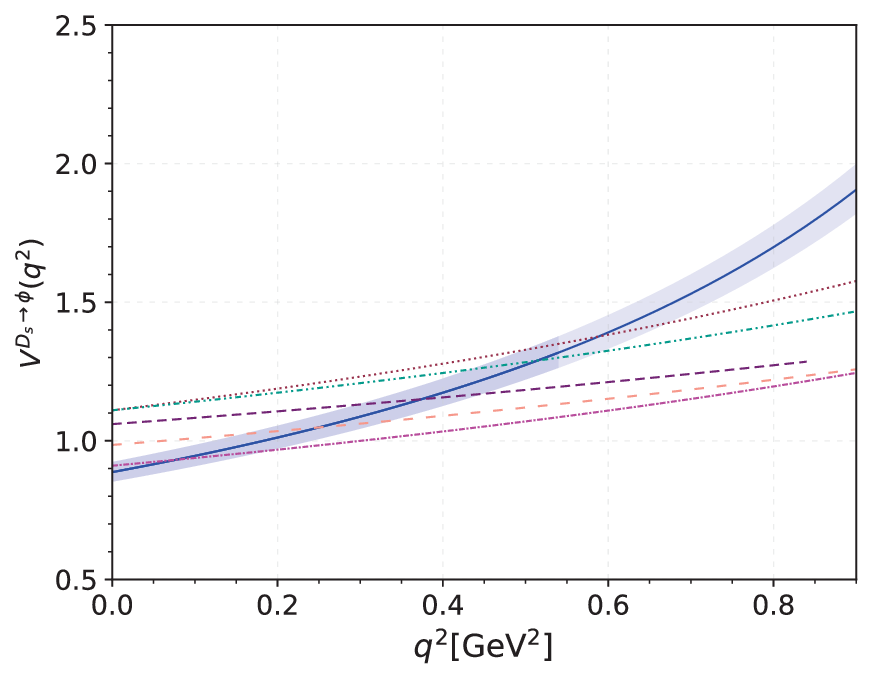}
\caption{Curves of the TFFs $A_1(q^2)$, $A_2(q^2)$, and $V_0(q^2)$ of semi-leptonic decays $D^0\to\rho^-\ell^+\nu_\ell$, $D^+\to \overline{K}^{\ast0}\ell^+\nu_\ell$, and $D_s^+\to\phi\ell^+\nu_\ell$ with $\ell=(e,\mu)$ versus $q^2$ in the entire physical region. The solid lines represent the central values, while the shaded bands indicate the corresponding errors. Predictions from other theoretical methods such as LCSRs~\cite{Lin:2025cmn}, CQM~\cite{Melikhov:2000yu}, LFQM~\cite{Verma:2011yw}, LQCD~\cite{Flynn:1997ca, Donald:2013pea}, HM$\chi$T~\cite{Fajfer:2005ug}, LEChQM~\cite{Palmer:2013yia}, RQM~\cite{Faustov:2019mqr}, CCQM~\cite{Ivanov:2019nqd} are also shown as a comparison.}
\label{fig:ttf0}
\end{figure*}

The LCSRs method is valid in the low and intermediate $q^2$ region. To obtain the decay widths and branching fractions of $D\to(\rho, K^\ast)$ and $D_s\to\phi$ semi-leptonic decays, the TFFs must be extrapolated to the full physical region, i.e., $q^2\in [0, (m_{D_{(s)}} - m_V)^2]$. This work employs the Simplified Series Expansion (SSE) method based on $z(t)$-expansion for extrapolation. This approach translates the threshold behavior of TFFs into constraints on expansion coefficients, while its rapid convergence properties combined with conventional pole model parameterization ensure the physical reliability of the extrapolated results~\cite{Khodjamirian:2010vf,Bharucha:2015bzk}. Then, the TFFs (without loss of generality, represented by $f$) take the following form,
\begin{align}
f_i(q^2)=\frac{1}{1-q^2/m^2_{R,i}}\sum_{k=0,1,2}a_k^i[z(q^2)-z(0)]^k,
\end{align}
The conformal map $z(q^2,t_0)$ is defined as
\begin{align}
z(q^2,t_0)=\frac{\sqrt{t_+-q^2}-\sqrt{t_+-t_0}}{\sqrt{t_+-q^2}+\sqrt{t_+-t_0}},
\end{align}
where $t_0=t_+(1-\sqrt{1-t_-/t_+})$ and $t_{\pm}=(m_{D_{(s)}}\pm m_V)^2$, and the resonance masses $m_{R,i}$ are selected based on the $J^P$ quantum numbers of $D_s^+$ and $D$ mesons. These mass parameters are determined with reference to theoretical recommendations from Refs.~\cite{Momeni:2020zrb, ParticleDataGroup:2024cfk}. To ensure the reliability of the extrapolated results, we optimize the free parameters $a_1^i$ and $a_2^i$ to minimize the evaluation parameter $\Delta_i$, and requiring $\Delta_i < 1\%$. This parameter is defined as the sum of relative deviations between theoretical values $f_i(t)$ and fitted values $f_i^{\mathrm{fit}}(t)$ at discrete points, $\Delta_i = \sum |f_i(t) - f_i^{\mathrm{fit}}(t)| / \sum |f_i(t)| \times 100\%$, where the squared momentum transfer $t\in[0,1/40,...,40/40]\times0.8\rm~GeV^2$ spans the interval $[0, (m_{D_{(s)}}-m_V)^2]~\text{GeV}^2$ with equidistant sampling points. Based on the LCSRs and the $z$-series expansion method, the extrapolated TFFs for $D_{(s)}\to V$ over the entire $q^2$ range are shown in Fig.~\ref{fig:ttf0}. The results of other theoretical methods such as LCSRs~\cite{Lin:2025cmn}, HM$\chi$T~\cite{Fajfer:2005ug}, covariant quark model (CQM)~\cite{Melikhov:2000yu}, CCQM~\cite{Ivanov:2019nqd}, LFQM (2011)~\cite{Verma:2011yw}, and LQCD~\cite{Donald:2013pea,Flynn:1997ca}, LEChQM~\cite{Palmer:2013yia}, RQM~\cite{Faustov:2019mqr} are also given in this figure for comparison. Our $A^{D\to\rho}_2(q^2)$ is consistent with the predictions of LFQM~\cite{Verma:2011yw}, LEChQM~\cite{Palmer:2013yia} in the whole error range. The $A^{D\to K^\ast}_2(q^2)$ by us is consistent with the results of CQM~\cite{Melikhov:2000yu} and RQM~\cite{Faustov:2019mqr}. Our $A^{D\to\phi}_2(q^2)$ is in good agreement with the predictions of LQCD~\cite{Flynn:1997ca} and HM$\chi$T~\cite{Fajfer:2005ug} in the intermediate and low $q^2$ regions. In addition, the behaviors of our $A_1^{D_{(s)}\to V}(q^2)$ and $V^{D_{(s)}\to V}(q^2)$ are steeper than that by other theoretical methods. In low $q^2$ region, however, our $V^{D\to \rho}(q^2)$ agrees with that by LCSRs~\cite{Lin:2025cmn} and LEChQM~\cite{Palmer:2013yia}, our $V^{D\to K^\ast}(q^2)$ is consistent with the one by CCQM~\cite{Ivanov:2019nqd}, and our $V^{D_s\to \phi}(q^2)$ is in agreement with that by LFQM~\cite{Verma:2011yw}, respectively.

By combining the decay width formula~\eqref{eq:GamL} and~\eqref{eq:GamT} with the CKM matrix elements $|V_{cs}|=0.975\pm0.006$ and $|V_{cd}|=0.221\pm0.004$~\cite{ParticleDataGroup:2024cfk}, integrating over the entire physical $q^2$ region, we obtain the total decay width and the longitudinal-to-transverse helicity decay width ratio as follows,
\begin{align}
&\Gamma_{\rm total}^{D\to\rho} = (61.424^{+8.858}_{-7.899})\times10^{-15}, \nonumber \\[4pt]
&\frac{\Gamma_L^{D\to\rho}}{\Gamma_T^{D\to\rho}} = 1.022^{+0.014}_{-0.015}, \nonumber \\[4pt]
&\Gamma_{\rm total}^{D\to K^\ast} = (37.288^{+4.751}_{-4.242})\times10^{-15}, \nonumber \\[4pt]
&\frac{\Gamma_L^{D\to K^\ast}}{\Gamma_T^{D\to K^\ast}} = 1.112^{+0.036}_{-0.021}, \nonumber \\[4pt]
&\Gamma_{\rm total}^{D_s\to\phi} =(34.457^{+4.515}_{-3.293})\times10^{-15}, \nonumber \\[4pt]
&\frac{\Gamma_L^{D_s\to\phi}}{\Gamma_T^{D_s\to\phi}} = 1.045^{+0.032}_{-0.037}.
\label{eq:Dec_Wid}
\end{align}
Our $\Gamma_L^{D_s\to\phi}/\Gamma_T^{D_s\to\phi}$ is consistent with the CLEO experimental data $\Gamma_L^{D_s\to\phi}/\Gamma_T^{D_s\to\phi} = 1.0 \pm 0.3 \pm 0.2$~\cite{CLEO:1994msc} within uncertainties.

Finally, by using the lifetime  $\tau_{D_s^+}=0.5012\pm0.0022$ ps, $\tau_{D^+}=1.033\pm0.005$ ps and $\tau_{D^0}=0.4103\pm0.001$ ps~\cite{ParticleDataGroup:2024cfk}, we calculate the branching  fractions for the semi-leptonic decay processes $D^0\to\rho^-\ell^+\nu_\ell$, $D^+\to \overline{K}^{\ast0}\ell^+\nu_\ell$, and $D_s^+\to\phi\ell^+\nu_\ell$ with $\ell=(e,\mu)$. Those predictions are listed in Tables~\ref{table:bf3},~\ref{table:bf2},~\ref{table:bf1}. For comparison, we also list the results from
BESIII~\cite{BESIII:2021pvy, BESIII:2018qmf, BESIII:2017ikf, BESIII:2023opt,
BESIII:2024lxg}, CLEO~\cite{CLEO:2011ab, CLEO:2005rxg, Hietala:2015jqa, CLEO:2010enr}, BaBar~\cite{BaBar:2008gpr}, PDG~\cite{ParticleDataGroup:2024cfk}, LQCD~\cite{APE:1994kxx}, LCSRs~\cite{Wu:2006rd,Leng:2020fei,Aliev:2004vf,Fu:2020vqd}, CQM~\cite{Melikhov:2000yu,Soni:2017eug}, LFQM~\cite{Cheng:2017pcq}, HM$\chi$T~\cite{Fajfer:2005ug}, RQM~\cite{Faustov:2019mqr}, CCQM~\cite{Ivanov:2019nqd,Soni:2018adu}, ISGW2~\cite{Scora:1995ty}, HQEFT~\cite{Wang:2002zba,Wu:2006rd}, chiral unitary approach ($\chi$UA)~\cite{Sekihara:2015iha}, CLFQM~\cite{Cheng:2017pcq,Wang:2008ci}, 3PSRs~\cite{Du:2003ja}, symmetry-preserving regularisation of a vector-vector contact interaction (SCI)~\cite{Xing:2022sor}. One can find that,
\begin{itemize}
\item For $\mathcal{B}(D^0\to\rho^-e^+\nu_e)$, the central value of our prediction lies within the error bands of the CLEO 2005~\cite{CLEO:2005rxg} and 2013~\cite{CLEO:2011ab} measurements. It agrees with the RQM~\cite{Faustov:2019mqr}, $\chi$UA~\cite{Sekihara:2015iha}, and LCSR~\cite{Wu:2006rd} results within $1\sigma$. For $\mathcal{B}(D^0\to\rho^-\mu^+\nu_{\mu})$, our prediction agrees well with the $\chi$UA~\cite{Sekihara:2015iha} result.
\item For $\mathcal{B}(D^+\to\overline{K}^{\ast0} e^+\nu_e)$, our central values are numerically close to the measured results of the CLEO~\cite{CLEO:2005rxg,CLEO:2010enr} and the PDG~\cite{ParticleDataGroup:2024cfk}, respectively. For $\mathcal{B}(D^+\to\overline{K}^{\ast0} \mu^+\nu_{\mu})$, our prediction agrees well with the LCSR~\cite{Fu:2020vqd,Wu:2006rd}, HM$\chi$T~\cite{Fajfer:2005ug}, and $\chi$UA
    ~\cite{Sekihara:2015iha} results, with no significant deviations.
\item For $\mathcal{B}(D_s^+\to\phi e^+\nu_e)$, our result is in agreement with the PDG~\cite{ParticleDataGroup:2024cfk}, the BESIII~\cite{BESIII:2017ikf} and CLEO~\cite{Hietala:2015jqa} measurements. For $\mathcal{B}(D_s^+\to\phi \mu^+\nu_\mu)$, our prediction is consistent with the latest BESIII measurement~\cite{BESIII:2023opt} within errors.
\end{itemize}

\begin{table}[htp]
\renewcommand{\arraystretch}{1.2}
\scalebox{0.2}
\footnotesize
\caption{Branching fractions of the semi-leptonic decays $D^0\to\rho^-\ell^+\nu_{\ell}$ (in units of $10^{-3}$) from our results and some different methods.}
\label{table:bf3}
\begin{tabular}{lllll}
\hline\hline
&$\mathcal{B}(D^0\to\rho^- e^+\nu_e)$&$\mathcal{B}\left(D^0\to\rho^-\mu^+\nu_\mu\right)$&  \\
\hline
This work                     & $1.877^{+0.271}_{-0.241}$       &$1.868^{+0.269}_{-0.240}$ \\[2pt]
BESIII'24~\cite{BESIII:2024lxg}    & $1.439\pm0.033\pm0.027$        &$-$ \\[2pt]
BESIII'21~\cite{BESIII:2021pvy}    & $-$                            &$1.35\pm0.09\pm0.09$ \\[2pt]
BESIII'18~\cite{BESIII:2018qmf}    & $1.445\pm0.058\pm0.039$        &$-$ \\[2pt]
CLEO'13~\cite{CLEO:2011ab}      & $1.77\pm0.12\pm0.10$           &$-$ \\[2pt]
CLEO'05~\cite{CLEO:2005rxg}      & $1.94\pm0.39\pm0.13$          &$-$ \\[2pt]
LFQM~\cite{Cheng:2017pcq}         & $-$                            &$1.7\pm0.2$ \\[2pt]
LCSRs~\cite{Wu:2006rd}             & $1.81^{+0.18}_{-0.13}$         &$1.73^{+0.17}_{-0.13}$ \\[2pt]
LCSRs~\cite{Leng:2020fei}          & $1.74\pm0.25$                    &$1.65\pm0.23$ \\[2pt]
RQM~\cite{Faustov:2019mqr}         & $1.96$                         &$1.88$ \\[2pt]
ISGW2~\cite{Scora:1995ty}           & $1.0$                         &$-$ \\[2pt]
HM$\chi$T~\cite{Fajfer:2005ug}            & $2.0$                         &$-$ \\[2pt]
HQEFT~\cite{Wang:2002zba}            & $1.4\pm0.3$                   &$-$ \\[2pt]
CCQM~\cite{Soni:2018adu}             & $1.62$                        &$1.55$ \\[2pt]
$\chi$UA~\cite{Sekihara:2015iha}          & $1.97$                        &$1.84$ \\[2pt]
\hline\hline
\end{tabular}
\end{table}
\begin{table}[hpt]
\renewcommand{\arraystretch}{1.2}
\scalebox{0.6}
\footnotesize
\caption{Branching fractions of the semi-leptonic decays $D^+\to \overline{K}^{\ast0}\ell^+\nu_\ell$ (in unit $10^{-2}$) from our results and some different methods.}
\label{table:bf2}
\begin{tabular}{lllll}
\hline\hline
&$\mathcal{B}(D^+\to \overline{K}^{\ast0} e^+\nu_e)$&$\mathcal{B}\left(D^+\to \overline{K}^{\ast0}\mu^+\nu_\mu\right)$&  \\
\hline
This work                             & $5.558^{+0.712}_{-0.635}$        &$5.546^{+0.707}_{-0.632}$ \\[2pt]
PDG~\cite{ParticleDataGroup:2024cfk}   & $5.40\pm0.10$                    &$5.26\pm0.15$ \\[2pt]
CLEO'05~\cite{CLEO:2005rxg}              & $5.56\pm0.27\pm0.23$             &$-$ \\[2pt]
CLEO'10~\cite{CLEO:2010enr}              & $5.52\pm0.07\pm0.13$             &$5.27\pm0.07\pm0.14$ \\[2pt]
LQCD~\cite{APE:1994kxx}                & $6.26\pm0.184$                    &$5.95\pm0.167$ \\[2pt] LCSRs~\cite{Fu:2020vqd}              & $5.282^{+0.847}_{-0.796}$        &$5.242^{+0.838}_{-0.787}$ \\[2pt]
LCSRs~\cite{Wu:2006rd}               & $5.37^{+0.24}_{-0.23}$           &$5.10^{+0.23}_{-0.21}$ \\[2pt]
CCQM~\cite{Ivanov:2019nqd}                & $7.61$                           &$7.21$ \\[2pt]
CQM~\cite{Soni:2017eug}                 & $8.35$                           &$7.94$ \\[2pt]
CQM~\cite{Melikhov:2000yu}              & $6.24$                           &$-$ \\[2pt]
$\chi$UA~\cite{Sekihara:2015iha}          & $5.56$                          &$5.12$ \\[2pt]
HM$\chi$T~\cite{Fajfer:2005ug}               & $5.6$                           &$5.6$ \\[2pt]
\hline\hline
\end{tabular}
\end{table}
\begin{table}[hpt]
\renewcommand{\arraystretch}{1.2}
\scalebox{0.6}
\footnotesize
\caption{Branching fractions of the semi-leptonic decays $D_s^+\to\phi\ell^+\nu_\ell$ (in unit $10^{-2}$) from our results and some different methods.}
\label{table:bf1}
\begin{tabular}{lllll}
\hline\hline
&$\mathcal{B}(D_s^+\to\phi e^+\nu_e)$&$\mathcal{B}\left(D_s^+\to\phi\mu^+\nu_\mu\right)$&  \\
\hline
This work                     &$2.501^{+0.329}_{-0.240}$            &$2.489^{+0.327}_{-0.239}$\\[2pt]
PDG~\cite{ParticleDataGroup:2024cfk}      & $2.34\pm0.12            $&$2.24\pm0.11$ \\
BESIII'17~\cite{BESIII:2017ikf}         & $2.26\pm0.45\pm0.09$         &$1.94\pm0.53\pm0.09$ \\[2pt]
BESIII'23~\cite{BESIII:2023opt}          & $-$                         &$2.25\pm0.09\pm0.07$ \\[2pt]
CLEO~\cite{Hietala:2015jqa}              & $2.14\pm0.17\pm0.08$        &$-$ \\[2pt]
BaBar~\cite{BaBar:2008gpr}               & $2.61\pm0.11\pm0.15$        &$-$ \\[2pt]
CCQM~\cite{Ivanov:2019nqd}     &$3.01$                               & $2.85$ \\[2pt]
CQM~\cite{Melikhov:2000yu}       &2.57                                & 2.57& \\[2pt]
3PSR~\cite{Du:2003ja}          &$1.80\pm0.50 $                     & $-$&\\[2pt]
CLFQM~\cite{Cheng:2017pcq}      &$3.1\pm0.3$                       & $2.9\pm0.3$&\\[2pt]
CLFQM~\cite{Wang:2008ci}        &2.30                              & $-$&\\[2pt]
LCSRs~\cite{Aliev:2004vf}           &$2.15^{+0.27}_{-0.31} $           & $-$&\\[2pt]
HQEFT~\cite{Wu:2006rd}             &$2.53^{+0.37}_{-0.40} $           & $2.40^{+0.35}_{-0.40}$& \\[2pt]
RQM~\cite{Faustov:2019mqr}         &$2.69 $                           & $-$& \\[2pt]
SCI~\cite{Xing:2022sor}            &$2.45 $                           & $2.30$& \\
\hline\hline
\end{tabular}
\end{table}

\section{Summary}\label{biao3}
In this paper, we study the leading-twist longitudinal DA of the vector $\rho$, $K^\ast$, and $\phi$ mesons. The study scheme suggested in Ref.~\cite{Zhong:2021epq} by us is adopted. The $\xi$-moments $\langle\xi^n\rangle_{2;V}^\parallel$ are calculated with QCD SRs in the framework of the BFT. Numerically, a new sum rule formula~\eqref{eq:xiSR_new} for $\langle\xi^n\rangle_{2;V}^\parallel$ based on the fact that the sum rule of $\langle\xi^0\rangle_{2;V}^\parallel$ cannot be normalized in the whole $M^2$ region is adopted. Our predictions for the first ten $\xi$-moments $\langle\xi^n\rangle_{2;V}^\parallel$ are exhibited in Tables~\ref{tab:xi_value1} and~\ref{tab:xi_value2}. With the simple PLP model, by fitting those resulted $\langle\xi^n\rangle_{2;V}^\parallel$ with the least squares method, the behaviors of DA $\phi_{2;V}^\parallel(x,\mu)$ are determined, and shown in Figs.~\ref{fig:DAs}. Our DA behaviors are very close to that by DSE for $\phi_{2;\rho}^\parallel(x,\mu)$, close to that by LQCD and Asymptotic form for $\phi_{2;K^\ast}^\parallel(x,\mu)$ but the peak deviates from $x = 0.5$ due to the $SU_f(3)$ breading effect, very close to that by DSE and LQCD for $\phi_{2;\phi}^\parallel(x,\mu)$. In addition, we also analyze the impact of incorporating different numbers of $\xi$-moments into the fitting on the stability and reliability of the obtained DAs. We find that only small changes in the fitted parameters and reconstructed DA curves as NS varies. The results indicate that the lowest few $\xi$-moments are sufficient to determine the DA shapes. However, we believe that it is still necessary to include more higher-order $\xi$-moments. The reason is that the higher-order moments provide additional non-perturbative constraints, and allow us to effectively test the stability and internal consistency of DA reconstruction.

Further, we use the obtained DA $\phi_{2;\rho}^\parallel(x,\mu)$, $\phi_{2;K^\ast}^\parallel(x,\mu)$ and $\phi_{2;\phi}^\parallel(x,\mu)$ to study the semi-leptonic decays $D\to\rho/K^\ast\ell^+\nu_\ell$ and $D_s\to\phi \ell^+\nu_\ell$ with LCSRs. The TFFs at the large recoil point ($q^2=0~\mathrm{GeV}^2$) are exhibited in Tables \ref{table:ttf1},~\ref{table:ttf2},~\ref{table:ttf3}. We also give two important TFFs ratios, $r_V$ and $r_2$. Our $r_V^{D\to \rho}$ and $r_V^{D_s\to\phi}$ values agree with the measurement of BESIII~\cite{BESIII:2024lxg} and BESIII~\cite{BESIII:2023opt} within errors, respectively. After extrapolating with the SSE method based on the $z(t)$-expansion, the behaviors of those TFFs in the whole $q^2$ region are obtained and shown in Fig.~\ref{fig:ttf0}. Finally, we calculate the branching fractions and decay widths for the semi-leptonic decays $D^0\to\rho^-\ell^+\nu_\ell$, $D^+\to \overline{K}^\ast\ell^+\nu_\ell$, and $D_s^+\to\phi\ell^+\nu_\ell$ with $\ell=(e,\mu)$. The results are summarized in Tables~\ref{table:bf3},~\ref{table:bf2},~\ref{table:bf1}.
We compare our predictions with the BESIII and CLEO measurements. They agree with each other within errors.

\section{Acknowledgments}
This work was supported in part by the National Natural Science Foundation of China under Grants No. 12265009, the Project of Guizhou Provincial Department of Science and Technology under Grant No. MS[2026]306.
\\
\appendix
\section{The rest of the condensation terms}\label{appendix}
The specific expressions of $\mathcal{O}(m^2)$ mass corrections for double-gluon condensate $\langle\alpha_s G^2\rangle$, triple-gluon condensate $\langle g_s^3fG^3 \rangle$ and four-quark condensate $\langle g_s^2\bar qq\rangle^2$ in Eq.~\eqref{eq:xiSR} are as follows.
\begin{widetext}
\begin{align}
  \hat{I}_{\langle G^2\rangle}(M^2)&=\frac{\langle\alpha_{s}G^{2}\rangle( m_{1}^2+(-1)^nm_2^2)}{72(M^2)^3 \pi}\,\bigg\{12\bigg[\psi(n+1)-\ln \frac{M^2}{\mu^2}+2\gamma_E\bigg]\!-\!20+\theta(n-1)\bigg[12n\bigg
  (\psi(n+1)-\ln \frac{M^2}{\mu^2}+2\gamma_E\bigg)
\nonumber\\[5pt]
&
   -3(-1)^n \psi_{4}(n)\, \bigg]+\theta (n-2)\bigg[(-1)^n \bigg((6n+3)\psi_{3}(n)-\frac{6(n+1)((-1)^nn+1)}
   {n}\bigg)\bigg]\bigg\},
\\
     \hat{I}_{\langle G^{3}\rangle}(M^{2})&=\frac{\langle g_{s}^{3}fG^{3}\rangle}{3456 (M^2)^4 \pi^2}\bigg\{(m_{1}^2+(-1)^n m_2^2)\bigg\{-33-9(-1)^n+\theta (n-1)\bigg[\,48(3n-1) n\bigg(\psi(n+1)-\,\ln\frac{M^2}{\mu^2}+2\gamma_E\bigg)
\nonumber\\[5pt]
&
  -270n^2+2n(-1)^n(9+47 (-1)^n) +96\bigg]+\theta(n-2)\bigg[6(-1)^n\bigg(63 (-1)^n n^2-53 (-1)^n n-6(-1)^n\! \! +4\bigg)\!\!-24n
\nonumber\\[5pt]
&
    \times(-1)^n\psi_{3}(n)\bigg]+\theta(n-3)\bigg[\frac{-18(-1)^n}{n-1}\bigg(\left(4\psi_{1}(n)+5 (-1)^n\right) n^3-4(\psi_{1}(n)+1+(-1)^n)n^2 -(2+(-1)^n) n
\nonumber\\[5pt]
&
     -2\bigg) \bigg] \bigg\}-24(1+(-1)^n) m_{1} m_{2} \bigg\},
\\
     \hat{I}_{\langle q^{4}\rangle}(M^{2})=&\frac{\langle g_{s}^{2}\bar{q}q\rangle^{2}}{5832\pi^2}\frac
     {2+\kappa^2}{(M^2)^4}\bigg\{((-1)^nm_{1}^2+m_2^2)\bigg\{\bigg[\psi(n+1)-\ln\frac{M^2}{\mu^2}+2\gamma_E
      \bigg]\bigg[54-474  (-1)^n+\theta(n-1)~\bigg[12\bigg(92(-1)^n n^2
\nonumber\\[5pt]
&
  -\left(51\!+\!7 (-1)^n\right) n+44(-1)^n\bigg)\bigg]\!+\!\theta(n-2)\bigg[\!-\!264 \bigg(2(-1)^n n^2\!-\!2(-1)^n n+1+\!(-1)^n\bigg) \bigg]\bigg]\!+\!\theta (n-1)\bigg[6 \bigg(59n
\nonumber\\[5pt]
&
   +51-\frac{9}{n}\bigg)-\frac{2(-1)^n}{n} \bigg(1780n^3-85n^2-233 n+27\bigg)\bigg]+\theta(n-2)\bigg[
   \frac{306 n^3-68 n^2+290 n-264}{(n-1) n}+\frac{2(-1)^n}{(n-1) n}
\nonumber\\[5pt]
&
  \times\bigg(1480 n^4-2471 n^3+642 n^2+349n-132 \bigg)+\psi_{3}(n)\bigg((-1)^n(27-306 n)\!-\!42n+27\bigg)
   \,\bigg]+\theta(n-3)\bigg[\frac{24}{n-1}\,
\nonumber\\[5pt]
&
    \times\bigg(12 n^2+17 n-5\bigg) \!-\!24\,(-1)^n\, \bigg(\,15\, n^2\,+\,14n\,-\,11\bigg)
   +\psi_{1}(n)\bigg(\,132 (-1)^n\,-\,12 \left(\,24 n^2\,+\,22 n\,-\,11\,\right)\,\bigg)\, \, \bigg]
\nonumber\\[5pt]
&
+540-412 (-1)^n\bigg\}\! +\!6(1+(-1)^n)m_{1}m_{2}\bigg[\psi_3(n)\theta(n-2)\!-\!\frac{2}{n}\theta(n-1)+4+2 \bigg(\psi(n+1)\!-\!\ln \frac{M^2}{\mu^2}+2\gamma_E\bigg)\bigg]\bigg\},
\end{align}
\end{widetext}
where
\begin{align}\label{A4}
&\psi_{1}(n)=\psi\left(\frac{n}{2}\right)-\psi\left(\frac{n-1}{2}\right)-(-1)^n\ln4, \nonumber \\
&\psi_{2}(n)=\psi\left(\frac{n-1}{2}\right)-\psi\left(\frac{n-2}{2}\right)+(-1)^{n}\ln4, \nonumber\\
&\psi_{3}(n)=\psi\left(\frac{n+1}{2}\right)-\psi\left(\frac{n}{2}\right)+(-1)^n\ln4, \nonumber \\
&\psi_{4}(n)=\psi\left(\frac{n+2}{2}\right)-\psi\left(\frac{n+1}{2}\right)-(-1)^{n}\ln 4.
\end{align}

\end{document}